\documentclass[%
 reprint,
superscriptaddress,
 amsmath,amssymb,
 aps,
]{revtex4-2}

\usepackage{graphicx}
\usepackage{dcolumn}
\usepackage{bm}
\usepackage{braket}
\usepackage{siunitx}

\begin{document}

\preprint{APS/123-QED}

\title{Hyperfine-interaction limits polarization entanglement of photons from semiconductor quantum dots}%

\author{Christian Schimpf}
\email{christian.schimpf@eng.ox.ac.uk}
\affiliation{
 Institute of Semiconductor and Solid State Physics, Johannes Kepler University Linz, 4040 Linz, Austria
}
\author{Francesco Basso Basset}
\affiliation{
 Department of Physics, Sapienza University of Rome, 00185 Rome, Italy
}
\author{Maximilian Aigner}
\affiliation{
 Institute of Semiconductor and Solid State Physics, Johannes Kepler University Linz, 4040 Linz, Austria
}
\author{Wolfgang Attenender}
\affiliation{
 Institute of Semiconductor and Solid State Physics, Johannes Kepler University Linz, 4040 Linz, Austria
}
\author{Laia  Gínes}
\affiliation{
    Department of Physics, Stockholm University, 106 91 Stockholm, Sweden
}
\author{Gabriel Undeutsch}
\affiliation{
 Institute of Semiconductor and Solid State Physics, Johannes Kepler University Linz, 4040 Linz, Austria
}
\author{Marcus Reindl}
\affiliation{
 Institute of Semiconductor and Solid State Physics, Johannes Kepler University Linz, 4040 Linz, Austria
}
\author{Daniel Huber}
\affiliation{
 Institute of Semiconductor and Solid State Physics, Johannes Kepler University Linz, 4040 Linz, Austria
}
\author{Dorian Gangloff}
\affiliation{
    Department of Engineering Science, University of Oxford, Oxford, OX1 3PJ, United Kingdom
}
\author{Evgeny A. Chekhovich}
\affiliation{
    Department of Physics and Astronomy, University of Sheffield, United Kingdom
}
\author{Christian Schneider}
\affiliation{
Institute of Physics, Carl von Ossietzky Universit\"{a}t Oldenburg, 26111 Oldenburg, Germany
}
\author{Sven H\"{o}fling}
\affiliation{
Julius-Maximilians-Universit\"{a}t W\"{u}rzburg, Physikalisches Institut, Lehrstuhl f\"{u}r Technische Physik, Am Hubland, 97074 W\"{u}rzburg, Germany 
}
\author{Ana Predojevi\'{c}}
\affiliation{
    Department of Physics, Stockholm University, 106 91 Stockholm, Sweden
}
\author{Rinaldo Trotta}
\affiliation{
 Department of Physics, Sapienza University of Rome, 00185 Rome, Italy
}
\author{Armando Rastelli}
\affiliation{
 Institute of Semiconductor and Solid State Physics, Johannes Kepler University Linz, 4040 Linz, Austria
}
\date{\today}

\begin{abstract}
Excitons in quantum dots are excellent sources of polarization-entangled photon pairs, but a quantitative understanding of their interaction with the nuclear spin bath is still missing. Here we investigate the role of hyperfine energy shifts using experimentally accessible parameters and derive an upper limit to the achievable entanglement fidelity. Our results are consistent with all available literature, indicate that spin-noise is often the dominant process limiting the entanglement in InGaAs quantum dots, and suggest routes to alleviate its effect.
\end{abstract}

\maketitle



\begin{figure}
\includegraphics[trim=2mm 0 -2mm 0]{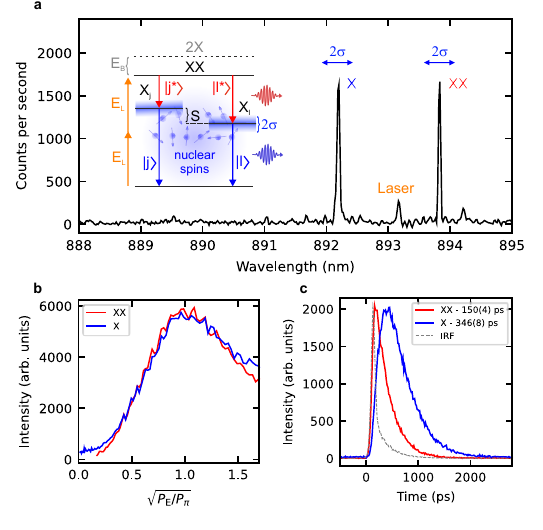}
\caption{\label{fig:pl} (a) Spectrum of an InGaAs QD (QD1) under two-photon-excitation (TPE). The sketch on the left depicts the biexciton-exciton (XX-X) level scheme. TPE is performed by tuning the energy $E_{\text{L}}$ of a pulsed laser to half of the XX energy. Here, the XX level lies $E_{\text{B}}=\SI{2.5}{meV}$ lower than two times the X energy (2X). The X states are split by the constant fine structure splitting $S$ and -- in the presence of randomly oriented nuclear spins -- also by the inhomogeneous Overhauser shift (OS) with a normally distributed amplitude with a standard deviation $\sigma$. (b) Rabi oscillation of the emitted XX and X signals with varying average laser power $P_\text{E}$ in multiples of the power $P_\pi\approx\SI{2}{\micro W}$ at the $\pi$-pulse condition. (c) Lifetime measurements of the XX and X emission. The depicted values are obtained by a convolved fit with the instruments response function (IRF).}
\end{figure}

Semiconductor quantum dots (QDs) have emerged as outstanding sources of single photons \cite{Senellart2017,Tomm2021} and photon pairs \cite{Liu2019,Wang2019}, allowing for high brightness and ultra-low multi-photon emission probability \cite{Schweickert2018,Neuwirth2022} for potential applications in quantum communication and information processing. Most of the proofs of principle demonstrations in these areas have relied on In(Ga)As QDs obtained via the Stranski-Krastanov (SK) epitaxial growth \cite{Vural2020} on GaAs substrates. 
A key advantage of QDs over other solid-state quantum emitters is their capability of emitting polarization-entangled photon pairs using the biexciton-exciton (XX-X) decay cascade \cite{Benson2000,Akopian2006,Young2006} [sketched in the inset of Fig \ref{fig:pl}(a)]. However, in spite of many careful investigations on SK and other In-containing QDs \cite{Akopian2006,Young2006,Hafenbrak2007,Salter2010,Dousse2010,Juska2013,Muller2014,Trotta2014,Olbrich2017,Mueller2018,Zeuner2021}, entanglement fidelity values above 0.9 could so far only be achieved by resorting to lossy time-filtering techniques \cite{Winik2017,Fognini2019,Anderson2020}. In contrast, fidelities values of up to 0.98 \cite{Huber2018} have been reported for GaAs QDs obtained via local Al-droplet etching (LDE) on AlGaAs \cite{Gurioli2019,DaSilva2021,Zhai2022}. The physical origin of this large discrepancy is still unclear and solving this puzzle would enhance the understanding of the physics of QDs in general and guide the improvement of the QD performance as sources of entangled photons. 

Several potential dephasing- and mixing-channels 
have been identified that could limit the polarization entanglement in QDs. Electron-hole exchange interaction causes the well-studied fine-structure-splitting (FSS) between the bright X levels~\cite{Bayer2002} in QDs with in-plane asymmetry. Other possible limitations mentioned in the literature are spin-scattering~\cite{Hudson2007}, phonon-induced dephasing~\cite{Hohenester2007,Reigue2017}, excitation-laser induced AC-Stark shifts~\cite{Seidelmann2022} and their combination~\cite{Seidelmann2023}. Nuclear spin noise in the form of stochastic hyperfine shifts, known as Overhauser fluctuations~\cite{Merkulov2002,Urbaszek2013,Kuhlmann2013}, have been proposed as the probable dominant dephasing mechanism in In(Ga)As QDs, motivated by the high nuclear spin (9/2) of In~\cite{Kuroda2013, Huber2017, keil2017} compared to Ga and As (3/2). However, no sufficiently accurate statements about the weight of nuclear spin noise have been made so far.

In this work we investigate the polarization entanglement of photon pairs generated by state-of-the-art InGaAs QDs \cite{He2013,Braun2016} with radiative lifetimes $T_1<\SI{500}{ps}$ and a very small FSS ranging from $\lesssim0.4$ to $\sim\SI{2.4}{\micro eV}$. We provide a quantitative estimate of the upper limit of the degree of entanglement achievable with such QDs due to nuclear spin noise and show that this limit is fully consistent with our measurements as well as former experimental results  \cite{Trotta2014,Hudson2007,zhang2015,Muller2014,Wang2019}, including those resorting to time-filtering \cite{Winik2017,Fognini2019,Zeuner2021,Anderson2020,Olbrich2017} and Purcell enhanced emission~\cite{Wang2019}. 
Our numerical results are based on  
direct measurements of the nuclear hyperfine-noise on the ground-state electron in InGaAs QDs, which reveal an inhomogeneous electron spin coherence time $T_2^*\simeq\SI{1.7}{ns}$ ~\cite{Urbaszek2013,Stockill2016,Gangloff2019}.
We further show that the consistently higher degree of entanglement found in LDE GaAs QDs \cite{Huber2018} can be tracked back to the lower $T_1$ ($\sim$\SI{230}{ps}) and the simultaneously higher measured electron spin $T_2^*$ ($\sim$\SI{2.6}{ns}~\cite{Zaporski2023}).


The energy levels relevant for generating entangled photon pairs from QDs are sketched in Fig. \ref{fig:pl}(a). The level structure of the neutral exciton states (X) is primarily defined by the electron-hole exchange-interaction \cite{Bayer2002}. An anisotropy of this interaction in the $x-y$ plane, perpendicular to the growth direction $z$, leads to a FSS between the ``bright" (optically dipole-allowed) excitons. In addition to the FSS, the interaction between the exciton and the nuclear spins leads to a hyperfine shift depending on the X spin configuration and nuclear spins´ orientations. The biexciton state (XX), as a singlet state, is neither affected by FSS nor by the hyperfine shifts. In the absence of stabilizing Knight-fields \cite{Urbaszek2013} or externally applied magnetic fields \cite{Bayer2002,Welander2014} the nuclear spins change their orientation and magnitude randomly at timescales in the order of $\tau_{\text{S}}\approx\SI{100}{\micro s}$ \cite{Merkulov2002,Kuhlmann2013}. Consequently, the OS changes many times during entanglement measurements, which take place on typical timescales of seconds, so the OS acts as an inhomogeneous broadening for the X level, following a Gaussian distribution.  The standard deviation $\sigma$ of the OS amplitudes can be identified via their connection to the inhomogeneous spin coherence time $T_2^*$ of the X. The hole's hyperfine contribution is typically at least 10 times smaller than the electron's~\cite{Eble2009}, so that the electron spin $T_2^*$ constitutes a good estimate of the X coherence time within about \SI{10}{\percent}. The expression for $\sigma$ then reads as~\cite{Merkulov2002,Stockill2016}:

\begin{equation}
\sigma \simeq \frac{\hbar}{T_2^*} = \sqrt{\frac{\sum_n x_n\,A_n\,I_n(I_n+1)}{N}},
\label{eq:sigma}
\end{equation}
with $x_n$ the fraction of the nuclear species $n$, $A_n$ the hyperfine coupling constant (in the order of \SI{50}{\micro eV} \cite{Stockill2016}), $I_n$ the nuclear spin and $N$ the number of nuclei in contact with the electron wavefunction. The electron spin $T_2^*$ can be determined by measuring its free induction decay via Ramsey interferometry~\cite{Ethier-Majcher2017,Gangloff2019}. For InGaAs QDs, values of around $T_2^*\simeq\SI{1.7}{ns}$ are typically observed~\cite{Press2010, Urbaszek2013, Stockill2016, Gangloff2019}, which results in $\sigma\simeq\SI{0.39}{\micro eV}$. From these values, a time-averaged loss of X coherence of approximately $1-\text{exp}[-(T_1/T_2^*)^2]=\SI{6}{\percent}$ \cite{Ethier-Majcher2017,Zaporski2023} can be expected, leading to a siginificant degradation of polarization entanglement. In comparison, for LDE GaAs QDs \cite{DaSilva2021}, with $T_1\simeq\SI{230}{ps}$ and $T_2^*\simeq\SI{2.6}{ns}$ \cite{Zaporski2023}, the loss amounts to $<\SI{1}{\percent}$, providing already a clear hint on the origin of the discrepancy between the fidelity values observed for SK InGaAs and LDE GaAs QDs.

We now investigate the generation of entangled photon pairs in individual InGaAs QDs at a temperature of \SI{5}{K}. The QDs were obtained via the Stranski-Krastanov growth mode and the partial capping and annealing method, resulting in $\sim$\SI{2}{nm}-high QDs with an average In fraction of $x_\text{In}=0.45(5)$ \cite{Braun2016}. The employed material is of state-of-the-art quality with high single-photon indistinguishability and an average X lifetime as low as \SI{400}{ps} (the samples originate from the same wafer as the one used in Ref.~\cite{He2013}). 
To limit the generation of free carriers in the QD surroundings, we create the XX state via coherent two-photon excitation (TPE)~\cite{Stufler2006}. The TPE process is sketched 
in the inset of Fig.~\ref{fig:pl}(a), where a pulsed laser with a pulse duration of \SI{7}{ps} is tuned to $E_{\text{L}}$, corresponding to half of the XX energy. To maximize chances of observing entanglement fidelities beyond state-of-the-art, we focus on QDs selected out of a much larger ensemble because of their intrinsically small value $S$ of the FSS, see inset of Fig. \ref{fig:pl}(a), which we obtain from polarization-resolved photoluminescence (PL) spectra with an accuracy of about \SI{0.2}{\micro eV} (see supplementary for details).
Figure \ref{fig:pl}(a) depicts the TPE-PL spectrum from an InGaAs QD (QD1), with $S=\SI{1.3(3)}{\micro eV}$ and an energy difference between the XX and the X photons of $E_{\text{B}}=\SI{2.4}{meV}$ (\SI{1.6}{nm}). Figure \ref{fig:pl}(b) shows the corresponding excitation dynamics under varying average laser power $P_\textrm{E}$. At the $\pi$-pulse condition (at the power $P_\pi\approx\SI{2}{\micro W}$), seen from the first Rabi-flop, the XX inversion efficiency is determined to be $\eta_\text{P}=0.70(2)$ (see supplementary for details). The radiative lifetimes for the XX-to-X and X-to-ground-state decays, obtained via a convolved fit of the traces in Fig.~\ref{fig:pl}(c) with the instrument response function (IRF), are $T_{1,\text{XX}}=\SI{150(4)}{ps}$ and  $T_{1,\text{X}}=\SI{346(8)}{ps}$, respectively. The $T_{1,\text{X}}$ value observed for this QD and others in the same sample is compatible with that observed under resonant excitation in Ref.~\cite{He2013} and favors the observation of highly entangled photons. The X (XX) lifetime depends strongly on the individual QD geometry \cite{Braun2016,Huber2019} and can therefore deviate from the average of about \SI{400}{ps} (\SI{200}{ps}) by up to \SI{30}{\percent}.

\begin{figure}
\includegraphics{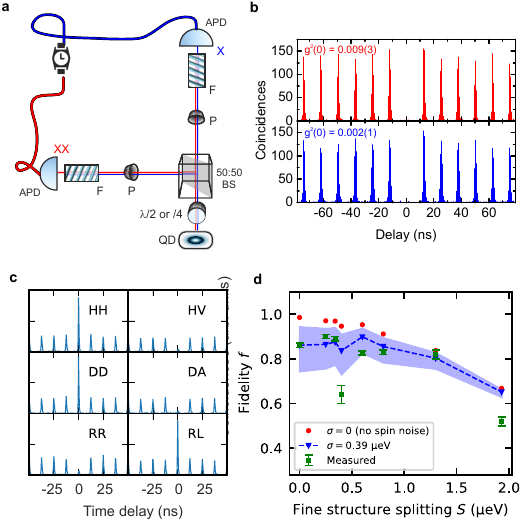}
\caption{\label{fig:fidelity} (a) Sketch of the experimental setup used to determine the entanglement fidelity of the QD photon source. A 50:50 non-polarizing beamsplitter (BS) is placed directly after the source. The exchangeable waveplate before and the polarizers (P) after the BS set the measurement basis of XX and X. The emitted photons are spectrally filtered (F) before impinging on avalanche photodiodes (APD), between which correlation measurements are performed. (b) HBT measurements performed on emitted XX (red) and X (blue) photons under pulsed, resonant TPE for QD2. The resulting $g^{(2)}(0)$ values, obtained by using a coincidence window of \SI{3}{ns}, are indicated in each panel. (c) Cross-correlation histograms for the six 2-qubit measurement bases used in the fidelity estimation for QD1p, with a resulting fidelity of 0.89(1). (d) Measured fidelity values for different QDs (squares) plotted against the respective FSS values. The red dots show the expected fidelity when only accounting for the FSS and the lifetime, the blue area depicts the range of expected values when including fluctuating Overhauser shifts (OS) according to the values of the electron $T_2^*$ reported in the literature.}
\end{figure}

The fidelity of a XX-X two-photon state $\rho$ to the $\ket{\phi^+}$ Bell state is defined as $f:=\braket{\phi^+|\rho|\phi^+}$. We estimate $f$ via the setup depicted in Fig. \ref{fig:fidelity}(a), in which the emitted state can be projected into arbitrary 2-qubit states by rotating the waveplate(s) in front of the beam-splitter (BS). When removing the polarizers, the same setup can be used to perform Hanbury Brown and Twiss measurements to obtain  auto-correlation histograms, as shown for QD2 in Fig.~\ref{fig:fidelity}(b). Using a coincidence window of \SI{3}{ns}, the $g^{(2)}(0)$ values for XX and X are found to be $g^{(2)}_{\text{XX}}(0)=0.009(3)$ and $g^{(2)}_{\text{X}}(0)=0.002(1)$, respectively. These values indicate a negligible multi-pair emission probability, comparable with former reports on similar QDs \cite{Muller2014}. The fidelity estimation is performed by inserting the polarizers and projecting the XX-X two-photon state into three bases, resulting in six cross-correlation histograms (Fig. \ref{fig:fidelity}(c) for QD1p, where ``p" stands for ``processed on a piezo-actuator", see supplementary for a full list of measured QDs). From these histograms a fidelity of $f=0.89(1)$ can be calculated (described in Ref. \cite{Hudson2007}). The fidelity measurement is repeated for seven different QDs on the same sample.
The resulting $f$ values are plotted (squares) as a function of $S$ in Fig. \ref{fig:fidelity}(d). In absence of dephasing mechanisms we would expect a monotonic decrease of the time-averaged $f$ with increasing $S$ beginning with $f\simeq 1$ at $S=0$, as a finite FSS induces a time-dependent evolution of the entangled state \cite{Hudson2007}. However, we find that none of the measured fidelities exceed a value of 0.9 and that reducing $S$ below \SI{1}{\micro eV} (corresponding to about the natural linewidth of the X transitions of $\delta_\text{F}\approx\SI{1.6}{\micro eV}$) leads to no significant increase in $f$. These findings, which are in line with previous reports, clearly indicate the presence of dephasing processes.

Considering the relevant properties of the studied QDs (X lifetime of about \SI{400}{ps}, corresponding to a natural linewidth $\delta_\text{F}$ of about \SI{1.6}{\micro eV}, which is much larger than the minimum value of $S$) and the excitation dynamics (TPE with pulse lengths of \SI{7}{ps}, much shorter than the XX lifetime of \SI{200}{ps}), we can safely conclude that excitation-induced dephasing \cite{Seidelmann2022} and re-excitation cannot explain the observed reduction in $f$. Both effects combined would account for less than 0.01 drop of fidelity. The measurement temperature of \SI{5}{K} is sufficiently low to exclude significant phonon-induced dephasing~\cite{Reigue2017,Hohenester2007,Seidelmann2023}, as this would be incompatible with the negligible change of $f$ we observed when varying the temperature from \SI{5}{K} to \SI{20}{K} for two different QDs (see supplementary). Both QDs exhibit significantly different linewidths ($\leq 12\,\delta_\text{F}$ and $\approx 44\,\delta_\text{F}$), typically caused by the specific charge noise the QDs are exposed to \cite{Kuhlmann2013,Schimpf2019}. The fact that the observed fidelity for both QDs is almost identical lets us also exclude charge noise as a dominant dephasing mechanism. We therefore ascribe the limited maximum value of entanglement to the interplay of finite FSS and nuclear spin noise, as supported by the model we derive in the following. 

The typical bright-dark exciton splitting of $\sim\SI{100}{\micro eV}$ \cite{Bayer2002,Urbaszek2013} is much larger than the $S$ and $\sigma$ of the QDs considered here (in the order of µeV). Therefore we can describe the X dynamics in the 2-dimensional bright exciton subspace in good approximation. In the basis given by X$_{H/V}=\{(\ket{+1}\pm\ket{-1})/\sqrt{2}\}$, where $\ket{\pm1}$ are the eigenstates of the total electron-hole angular momentum along the $z$-direction, the Hamiltonian is then given by \cite{Bayer2002}:
\begin{equation}
    H= \begin{pmatrix}
   \delta_1 & \delta_2\\
   \delta_2^* & -\delta_1,
    \end{pmatrix}
\label{eq:H}
\end{equation}

\noindent where $\delta_1=S/2$ and $\delta_2=i h_z$, with $h_z$ the OS due to the finite nuclear spin polarization in the QD. In the frozen spin approximation (when $\tau_{\text{S}}\gg T_1$) \cite{Merkulov2002,Urbaszek2013} $H$ can be treated as time-independent and we can construct the entangled state of the two photons emitted in the XX-X radiative decay cascade for a fixed $h_z$ as
\begin{equation}
    \ket{\psi_{h_z}}(t) = \frac{1}{\sqrt{2}} \left( \ket{j^*} \otimes \ket{j} + e^{-\frac{i}{\hbar} \, \Delta_E \, t} \ket{l^*} \otimes \ket{l} \right),
\label{eq:psi}
\end{equation}
with $t$ the emission time of the X photon relative to that of the XX photon. The orthonormal eigenstates $\ket{j}$ and $\ket{l}$ of Eq. \eqref{eq:H} are defined in the basis of the horizontal $\ket{H}$ and vertical $\ket{V}$ polarization and have an energy difference $\Delta_E= 2\sqrt{\delta_1^2+\delta_2^2}$. Due to conservation of angular momentum, the XX photon eigenstates are necessarily the complex conjugates ($^*$) of the X photon eigenstates (see supplementary for details). The time evolution of the two-photon density matrix $\rho_{h_z}(t)=\ket{\psi_{h_z}}\bra{\psi_{h_z}}$ is then given by
\begin{equation}
    i\,\hbar\,\frac{\text{d}}{\text{d}t}\rho_{h_z}(t) = \left[\mathbb{I}^{(2)} \otimes {H}, \rho_{h_z}(t) \right],
\label{eq:dtrho}
\end{equation}
with $\mathbb{I}^{(2)}$ the $2 \times 2$ identity matrix.
The differential equation Eq.~\eqref{eq:dtrho} can be readily solved numerically. The time-averaged density matrix $\braket{\rho}_t$ is calculated by drawing $h_z$ multiple times from a normal distribution with a mean value of zero and a standard deviation of $\sigma$. The value of $S$ remains constant. By indicating with $k$ the fraction of events with no more than one photon pair per excitation cycle, we account for the multi-photon emission probability $1-k$ as in \cite{Neuwirth2022}:

\begin{equation}
1-k\approx\frac{g^{(2)}_{\text{XX}}(0)+g^{(2)}_{\text{X}}(0)}{2}\,\eta_\text{P}.
\end{equation}
By measuring the quantities in the above equation for each QD, we add a mixing channel to $\braket{\rho}_t$ so that
\begin{equation}
    \tilde{\rho} = k\,\braket{\rho}_t + \frac{1-k}{4}\,\mathbb{I}^{(4)},
\label{eq:rhok}
\end{equation}
with $\mathbb{I}^{(4)}$ the $4\times 4$ identity matrix. In the supplementary we provide further details about the calculation of $\tilde{\rho}$ and how to derive the fidelity 
\begin{equation}
f = \frac{1}{4}\left(1 + k + \frac{2k}{1+4\,T_1^2\left(S^2+\sigma^2 \right)/\hbar^2} \right).
\label{eq:fid}
\end{equation}
Previously derived equations take spin-noise into account only via a phenomenological spin-scattering \cite{Hudson2007}, which acts as a mixing channel. However, the physical origin of such effects has not been sufficiently clarified and no clear connection to measured quantities have been drawn so far. The stochastic energy shift $\sigma$ in our model, however, manifests as a pure dephasing channel and is a direct consequence of the random OS measured directly in experiments on electron spins confined in QDs~\cite{Merkulov2002,Press2010,Urbaszek2013,Ethier-Majcher2017,Gangloff2019}. 

The red dots in Fig.~\ref{fig:fidelity}(d), which depict the fidelity calculated from Eq.~\eqref{eq:fid} with $\sigma=0$ and using the measured $S$,  $T_{1}$ and $k$ for each individual dot,  confirm that in the absence of other mechanisms, $f$ should reach values exceeding 0.98. The blue points correspond instead to the estimation for $\sigma\simeq\SI{0.39}{\micro eV}$ ($T_2^*\simeq\SI{1.7}{ns}$ \cite{Press2010,Gangloff2019}) and the filled blue area depicts the range of $f$ between the shortest ($T_2^*=\SI{1}{ns}$ \cite{Urbaszek2013}) and longest ($T_2^*=\SI{3.2}{ns}$ \cite{Ethier-Majcher2017}) coherence times in In(Ga)As QDs reported in the literature. We see that no data point lies above the expectations from the model including the realistic effect of OS. Most importantly, our model reproduces the plateau of $f$ values for $S \lesssim \sigma$ and allows us to ascribe it to the dominant role played by the Overhauser shifts while, for $S \gg \sigma$ the effect of the FSS dominates.


\begin{figure}
\includegraphics{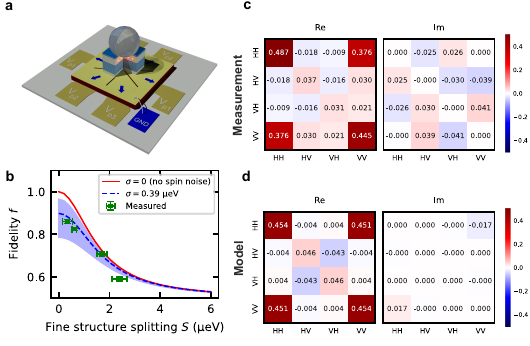}
\caption{\label{fig:sixleg} (a) Illustration of the piezoelectric device used to tune the FSS via three-axial strain-fields acting on QD1p, individually controlled via three input voltages $\left(V_{p1}, V_{p2}, V_{p3}\right)$ (b) Measured fidelity values of QD1p, for which the FSS was tuned to different values by the piezo-actuator. The red line shows the expected fidelity for $\sigma=0$ (no spin noise), the blue area depicts the expected fidelity range when including OS. (c) Real- and imaginary part of the 2-qubit density matrix, measured by state tomography on QD1p ($T_1=\SI{430(4)}{ps}$, $S=\SI{0.4(2)}{\micro eV}$ and $k=0.990(5)$) with 16 measurement bases, calculated using the maximum-likelihood approach with a resulting fidelity of $0.85(1)$, a purity of $0.74(2)$ and a concurrence of $0.69(2)$. (d) Modelled density matrix for QD1p, assuming $\sigma=\SI{0.41}{\micro eV}$ ($T_2^*=\SI{1.6}{ns}$), with a resulting fidelity of 0.89, a purity of 0.81 and a concurrence of 0.79.}
\end{figure}

Since the spin-properties of InGaAs QDs can vary strongly from QD to QD due to atomistic disorder related to alloying and consequent inhomogenous strain and piezoelectric fields, it is desirable to measure the FSS-dependent fidelity for one single QD (QD1p). Such measurements have been formerly conducted on InGaAs QDs using magnetic, electric, and/or strain fields \cite{Hudson2007, bennett2010,Trotta2014,zhang2015,lettner2021} but only under non-resonant excitation, which may introduce additional entanglement-degrading effects. To eliminate this uncertainty we use TPE and tune $S$ using a microprocessed piezoelectric strain-tuning actuator, as illustrated in Fig.~\ref{fig:sixleg}(a). This device allows the in-plane stress tensor in the QD structure to be manipulated by independently expanding or compressing three pairs of opposing ``legs" through the application of different voltages $\mathbf{V}=\left(V_{p1}, V_{p2}, V_{p3}\right)$ between the bottom contacts and the common ground-contact on top. The QD sample glued on the actuator was previously thinned down mechanically from 300 to \SI{30}{\micro m}, which still allows us to tune the FSS in a range of about \SI{5}{\micro eV}. The same kind of device was used in Ref. \cite{Huber2018}, where a fidelity of \SI{98}{\percent} was observed for LDE GaAs QDs. 

The green squares in Fig. \ref{fig:sixleg}(b) show the measured values of $f$ for different $S$, adjusted by varying $\mathbf{V}$. The red line shows the expected valued of $f$ according to Eq. \eqref{eq:fid} for $\sigma=0$. The blue dashed line, just like in the previous measurements, depicts the case for $\sigma=\SI{0.39}{\micro eV}$ ($T_2^*=\SI{1.7}{ns}$ \cite{Press2010, Gangloff2019}) and the filled blue area depicts the range of $f$ for the range of reported $T_2^*$ values. Again, all $f$ values lie well within the range predicted by Eq. \eqref{eq:fid}, when considering the measured $T_1$, $k$ and varied $S$ of QD1p.

To confirm the reliability of our fidelity measurements, which may be affected by polarization-altering effects in the used setup (Fig.~\ref{fig:fidelity}(a)),  we performed a state tomography with 16 measurement bases on QD1p at the lowest value of $S=\SI{0.4(2)}{\micro eV}$, obtaining the density matrix $\rho_\text{T}$ shown in Fig.~\ref{fig:sixleg}(c) via the maximum-likelihood estimator \cite{James2001}. The fidelity calculated from $\rho_\text{T}$ is $f=0.85(1)$, the purity and the concurrence $C$ \cite{Wootters1998}, which are independent from a unitary rotation of the state, are calculated as 0.74(2) and 0.69(2), respectively. The slightly lower $f$ compared to the 0.89(1) obtained by the fidelity estimation could stem from the additional wave-plates required for the state tomography, which are known to induce slight state-mixing by the inhomogeneities in the birefringent material. For comparison, the density matrix from Eq.~\eqref{eq:rhok} is shown in  Fig.~\ref{fig:sixleg}(d) using the measured values of $S$, $T_1$, and $k$. The value for the electron spin $T_2^*$ was estimated to $\SI{1.6}{ns}$ ($\sigma=\SI{0.41}{\micro eV}$) to match the $f=0.89$ measured using the fidelity estimation setup.
%
%

Before comparing the results of our model with former experimental results we emphasize that our model provides an upper limit for the degree of entanglement which may be observed experimentally, given the measured electron spin $T_2^*$. Other entanglement-degrading effects may still be present. As an example, the concurrence of 0.79 derived from the density matrix in Fig.~\ref{fig:sixleg}(d) is higher than the measured 0.69(2) and some entries of the experimentally reconstructed density matrix are not reproduced by the model (e.g. the $\braket{HH|\rho|HV}$ component).  
In addition, one data point in Fig.~\ref{fig:fidelity}, with $f\simeq 0.64$ for an $S$ of only \SI{0.4(2)}{\micro eV} deviates from the trend obtained from the others and was found to correspond to a QD showing a strong degree of linear polarization of the X signal, see supplementary.
This brings us to the conclusion that effects like heavy-hole--light-hole mixing \cite{Belhadj2010,tonin2012,Plumhof2013} or strain-activated quadrupolar double-spin-flips \cite{Bulutay2012,Urbaszek2013,Denning2019} could act as non-negligible mixing channels in InGaAs QDs (and possibly also in other material systems). In fact, state mixing could explain the discrepancy between the expected and the measured concurrence for the QD1p in Fig~\ref{fig:sixleg}, while the fidelity matches better. The reason is that the concurrence drops significantly faster compared to the fidelity in the presence of a mixing channel than for pure dephasing channels (see supplementary for details).

\begin{figure}
\includegraphics[trim={4mm 0cm -4mm 0cm}]{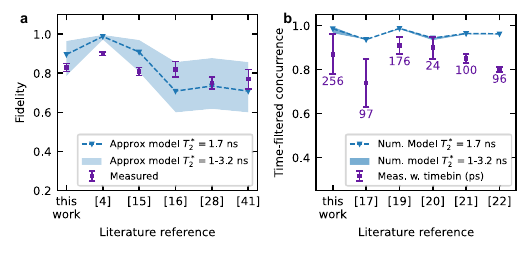}

\caption{\label{fig:comp} (a) Measured maximum $f$ from InGaAs QDs with $S\approx 0$ from the indicated references, compared against the values expected from Eq.~\eqref{eq:fid}, taking into account the reported $T_1$. The leftmost data point refers to QD1p. (b) Measured concurrence from the references indicated on the x axis, The measurements are compared against the values calculated by numerically propagating Eq. \eqref{eq:dtrho} from 0 up to the used coincidence window (indicated in the graph in ps), taking into account the reported values for $T_1$ and $S$. The leftmost datapoint refers to QD1p at its lowest value of $S$.}
\end{figure}

In Fig. \ref{fig:comp}(a) we show a compilation of $f$ values for InGaAs QDs with $S\leq\SI{0.4}{\micro eV}$ from representative work~\cite{Trotta2014,Hudson2007,zhang2015,Muller2014}, with $T_1$ ranging from \SI{400}{ps} to \SI{1}{ns} and one QD in a Purcell-enhancing circular Bragg resonators (CBRs) with $T_1=\SI{127(1)}{ps}$ and $S\leq \SI{1.2}{\micro eV}$ \cite{Wang2019} and compare them against the range of $f$ predicted by Eq. \eqref{eq:fid}. We see that the values consistently lie inside the range expected from the reported $T_2^*$ as used in Figs. \ref{fig:fidelity} and \ref{fig:sixleg}, indicating that OS are probably the dominant source of dephasing in those measurements. The CBR structure \cite{Wang2019} suffers from an AC Stark shift induced by the excitation laser, which becomes striking at low lifetimes \cite{Seidelmann2022}.
Finally, Fig.~\ref{fig:comp}(b) shows the values of $C$ from the literature, measured for other In-containing QDs with finite $S$ when relying on time-filtering \cite{Winik2017,Fognini2019,Zeuner2021,Anderson2020,Olbrich2017}. The emission wavelengths range from \SI{870}{nm} to \SI{1550}{nm}. The data points show the maximum achieved $C$ when including only photons within the coincidence windows stated in the graph. The values are compared against the range calculated numerically from Eq.~\eqref{eq:dtrho} again using the $T_2^*$ available only for InGaAs QDs, taking into account the coincidence window, $T_1$ and $S$ (see supplementary for details). We see that all measured values lie well below the maximum expected for the used coincidence window. The value of $C$ for QD1p (Fig. \ref{fig:sixleg}), increases from 0.69(2) to 0.87(9) when decreasing the coincidence window from \SI{3}{ns} to \SI{256}{ps}. Propagating Eq. \eqref{eq:dtrho} up to the detector resolution of \SI{350}{ps}, however, results in $C\approx 0.98$, again supporting the presence of additional entanglement-degrading processes requiring further attention. 


In summary, we have studied the behaviour of the polarization entanglement fidelity on several InGaAs quantum dots (QDs) with different fine structure splittings (FSS) and on a single strain-tuned QD on a sample bonded onto a piezoelectric actuator. For the latter, full state tomography was performed for the lowest achieved FSS to quantify the purity and the concurrence of the entangled state. The results are in good agreement with a theoretical model, which predicts an upper limit for the observable fidelity using the measured FSS of the exciton (X), its radiative lifetime, the multi-photon-pair emission probability, and the inhomogeneous electron coherence time $T_2^*$ as an input. From the model we conclude that slowly varying random Overhauser shifts (OS), prominent in QDs with abundant non-zero nuclear spins, lead to a pronounced dephasing of the bright X states over time, which limits the time-averaged polarization entanglement in InGaAs QDs. By comparing the model predictions with our data and values reported in the literature, we find that OS often constitute the dominant degradation mechanism. This result is consistent with the higher fidelity values of about 0.98 observed in LDE GaAs QDs, mainly due to the lower X lifetimes of about \SI{230}{ps}~\cite{Huber2018}. Additionally, the electron spin $T_2^*$ in LDE GaAs QDs was found to be higher ($\approx\SI{2.6}{ns}$) \cite{Zaporski2023}, as expected from Eq. \eqref{eq:sigma} when considering the lack of In and the approximately doubled number of nuclei $N$ in contact with the electron \cite{Zaporski2023}, compared to InGaAs QDs \cite{Gangloff2019}. Our results are also compatible with previous claims of dephasing-free In-containg QD sources of polarization-entangled photon pairs~\cite{Winik2017,Fognini2019} in the limit of low detector time-jitter $<\SI{50}{ps}$. In fact, the slowly varying OS, with correlation times above \SI{100}{\micro s}, lead to coherent phase rotations over the time $t$ elapsed between the XX and X recombination. This may lead to the observation of almost optimal entanglement for $t\ll T_2^*$ and entanglement degradation for $t$ comparable to $T_2^*$ due to averaging of the phase evolution occurring at different frequencies and directions. 
To improve the performance of InGaAs QDs as sources of entangled photons, the amplitude of the hyperfine fluctuations can be decreased by lowering the In-contribution or by developing QD structures in which the number of nuclei in contact with the electron wavefunction is increased and/or the X lifetime is decreased \cite{Huber2019}. Alternatively, the Overhauser fluctuations could be decreased by optically pre-cooling the nuclear spin bath \cite{Ethier-Majcher2017,Gangloff2019}, which provides a tuning knob for $\sigma$. These measures could boost the polarization entanglement of InGaAs QDs, making them compatible with state-of-the-art quantum communication protocols and demanding tasks in  quantum computation.

\medskip

The authors thank S. Moser for his assistance with the optical setup and U. Kainz for the help in device processing. This work has been supported by the Austrian Academy of Sciences (\"{O}AW), the QuantERA HYPER-U-P-S project, which received funding from the QuantERA ERA-NET Cofund in Quantum Technologies implemented within the European Union's
Horizon 2020 Programme and the Austrian Science Fund (FWF) I 3762, the European Union’s Horizon 2020 research and innovation program under Grant Agreements  No. 899814 (Qurope) and No. 871130 (Ascent+), the PNRR MUR project PE0000023-NQSTI (the National Quantum Science and Technology Institute)", the FFG Grant No. 891366, the Linz Institute of Technology (LIT) and the LIT Secure and Correct Systems Lab funded by the State of Upper Austria. We further acknowledge funding by the DFG within the project PR1749/1-1. A.P would like to acknowledge the support of the Swedish research council. L. G. was supported by the Knut\&Alice Wallenberg Foundation [through the Wallenberg Centre for Quantum Technology (WACQT)]. D.G. was supported by a University Research Fellowship from the Royal Society. E.A.C. was supported by the Royal Society and through an EPSRC grant EP/V048333/1.

\bibliography{main}

\providecommand{\noopsort}[1]{}\providecommand{\singleletter}[1]{#1}%
\begin{thebibliography}{60}%
\makeatletter
\providecommand \@ifxundefined [1]{%
 \@ifx{#1\undefined}
}%
\providecommand \@ifnum [1]{%
 \ifnum #1\expandafter \@firstoftwo
 \else \expandafter \@secondoftwo
 \fi
}%
\providecommand \@ifx [1]{%
 \ifx #1\expandafter \@firstoftwo
 \else \expandafter \@secondoftwo
 \fi
}%
\providecommand \natexlab [1]{#1}%
\providecommand \enquote  [1]{``#1''}%
\providecommand \bibnamefont  [1]{#1}%
\providecommand \bibfnamefont [1]{#1}%
\providecommand \citenamefont [1]{#1}%
\providecommand \href@noop [0]{\@secondoftwo}%
\providecommand \href [0]{\begingroup \@sanitize@url \@href}%
\providecommand \@href[1]{\@@startlink{#1}\@@href}%
\providecommand \@@href[1]{\endgroup#1\@@endlink}%
\providecommand \@sanitize@url [0]{\catcode `\\12\catcode `\$12\catcode
  `\&12\catcode `\#12\catcode `\^12\catcode `\_12\catcode `\%12\relax}%
\providecommand \@@startlink[1]{}%
\providecommand \@@endlink[0]{}%
\providecommand \url  [0]{\begingroup\@sanitize@url \@url }%
\providecommand \@url [1]{\endgroup\@href {#1}{\urlprefix }}%
\providecommand \urlprefix  [0]{URL }%
\providecommand \Eprint [0]{\href }%
\providecommand \doibase [0]{https://doi.org/}%
\providecommand \selectlanguage [0]{\@gobble}%
\providecommand \bibinfo  [0]{\@secondoftwo}%
\providecommand \bibfield  [0]{\@secondoftwo}%
\providecommand \translation [1]{[#1]}%
\providecommand \BibitemOpen [0]{}%
\providecommand \bibitemStop [0]{}%
\providecommand \bibitemNoStop [0]{.\EOS\space}%
\providecommand \EOS [0]{\spacefactor3000\relax}%
\providecommand \BibitemShut  [1]{\csname bibitem#1\endcsname}%
\let\auto@bib@innerbib\@empty
\bibitem [{\citenamefont {Senellart}\ \emph {et~al.}(2017)\citenamefont
  {Senellart}, \citenamefont {Solomon},\ and\ \citenamefont
  {White}}]{Senellart2017}%
  \BibitemOpen
  \bibfield  {author} {\bibinfo {author} {\bibfnamefont {P.}~\bibnamefont
  {Senellart}}, \bibinfo {author} {\bibfnamefont {G.}~\bibnamefont {Solomon}},\
  and\ \bibinfo {author} {\bibfnamefont {A.}~\bibnamefont {White}},\ }\bibfield
   {title} {\bibinfo {title} {{High-performance semiconductor quantum-dot
  single-photon sources}},\ }\href {https://doi.org/10.1038/nnano.2017.218}
  {\bibfield  {journal} {\bibinfo  {journal} {Nat. Nanotechnol.}\ }\textbf
  {\bibinfo {volume} {12}},\ \bibinfo {pages} {1026} (\bibinfo {year}
  {2017})}\BibitemShut {NoStop}%
\bibitem [{\citenamefont {Tomm}\ \emph {et~al.}(2021)\citenamefont {Tomm},
  \citenamefont {Javadi}, \citenamefont {Antoniadis}, \citenamefont {Najer},
  \citenamefont {L{\"{o}}bl}, \citenamefont {Korsch}, \citenamefont {Schott},
  \citenamefont {Valentin}, \citenamefont {Wieck}, \citenamefont {Ludwig},\
  and\ \citenamefont {Warburton}}]{Tomm2021}%
  \BibitemOpen
  \bibfield  {author} {\bibinfo {author} {\bibfnamefont {N.}~\bibnamefont
  {Tomm}}, \bibinfo {author} {\bibfnamefont {A.}~\bibnamefont {Javadi}},
  \bibinfo {author} {\bibfnamefont {N.~O.}\ \bibnamefont {Antoniadis}},
  \bibinfo {author} {\bibfnamefont {D.}~\bibnamefont {Najer}}, \bibinfo
  {author} {\bibfnamefont {M.~C.}\ \bibnamefont {L{\"{o}}bl}}, \bibinfo
  {author} {\bibfnamefont {A.~R.}\ \bibnamefont {Korsch}}, \bibinfo {author}
  {\bibfnamefont {R.}~\bibnamefont {Schott}}, \bibinfo {author} {\bibfnamefont
  {S.~R.}\ \bibnamefont {Valentin}}, \bibinfo {author} {\bibfnamefont {A.~D.}\
  \bibnamefont {Wieck}}, \bibinfo {author} {\bibfnamefont {A.}~\bibnamefont
  {Ludwig}},\ and\ \bibinfo {author} {\bibfnamefont {R.~J.}\ \bibnamefont
  {Warburton}},\ }\bibfield  {title} {\bibinfo {title} {{A bright and fast
  source of coherent single photons}},\ }\href
  {https://doi.org/10.1038/s41565-020-00831-x} {\bibfield  {journal} {\bibinfo
  {journal} {Nature Nanotechnology 2021 16:4}\ }\textbf {\bibinfo {volume}
  {16}},\ \bibinfo {pages} {399} (\bibinfo {year} {2021})}\BibitemShut
  {NoStop}%
\bibitem [{\citenamefont {Liu}\ \emph {et~al.}(2019)\citenamefont {Liu},
  \citenamefont {Su}, \citenamefont {Wei}, \citenamefont {Yao}, \citenamefont
  {da~Silva}, \citenamefont {Yu}, \citenamefont {Iles-Smith}, \citenamefont
  {Srinivasan}, \citenamefont {Rastelli}, \citenamefont {Li},\ and\
  \citenamefont {Wang}}]{Liu2019}%
  \BibitemOpen
  \bibfield  {author} {\bibinfo {author} {\bibfnamefont {J.}~\bibnamefont
  {Liu}}, \bibinfo {author} {\bibfnamefont {R.}~\bibnamefont {Su}}, \bibinfo
  {author} {\bibfnamefont {Y.}~\bibnamefont {Wei}}, \bibinfo {author}
  {\bibfnamefont {B.}~\bibnamefont {Yao}}, \bibinfo {author} {\bibfnamefont
  {S.~F.~C.}\ \bibnamefont {da~Silva}}, \bibinfo {author} {\bibfnamefont
  {Y.}~\bibnamefont {Yu}}, \bibinfo {author} {\bibfnamefont {J.}~\bibnamefont
  {Iles-Smith}}, \bibinfo {author} {\bibfnamefont {K.}~\bibnamefont
  {Srinivasan}}, \bibinfo {author} {\bibfnamefont {A.}~\bibnamefont
  {Rastelli}}, \bibinfo {author} {\bibfnamefont {J.}~\bibnamefont {Li}},\ and\
  \bibinfo {author} {\bibfnamefont {X.}~\bibnamefont {Wang}},\ }\bibfield
  {title} {\bibinfo {title} {{A solid-state source of strongly entangled photon
  pairs with high brightness and indistinguishability}},\ }\href
  {https://doi.org/10.1038/s41565-019-0435-9} {\bibfield  {journal} {\bibinfo
  {journal} {Nat. Nanotechnol.}\ }\textbf {\bibinfo {volume} {14}},\ \bibinfo
  {pages} {586} (\bibinfo {year} {2019})}\BibitemShut {NoStop}%
\bibitem [{\citenamefont {Wang}\ \emph {et~al.}(2019)\citenamefont {Wang},
  \citenamefont {Hu}, \citenamefont {Chung}, \citenamefont {Qin}, \citenamefont
  {Yang}, \citenamefont {Li}, \citenamefont {Liu}, \citenamefont {Zhong},
  \citenamefont {He}, \citenamefont {Ding}, \citenamefont {Deng}, \citenamefont
  {Dai}, \citenamefont {Huo}, \citenamefont {H{\"{o}}fling}, \citenamefont
  {Lu},\ and\ \citenamefont {Pan}}]{Wang2019}%
  \BibitemOpen
  \bibfield  {author} {\bibinfo {author} {\bibfnamefont {H.}~\bibnamefont
  {Wang}}, \bibinfo {author} {\bibfnamefont {H.}~\bibnamefont {Hu}}, \bibinfo
  {author} {\bibfnamefont {T.~H.}\ \bibnamefont {Chung}}, \bibinfo {author}
  {\bibfnamefont {J.}~\bibnamefont {Qin}}, \bibinfo {author} {\bibfnamefont
  {X.}~\bibnamefont {Yang}}, \bibinfo {author} {\bibfnamefont {J.~P.}\
  \bibnamefont {Li}}, \bibinfo {author} {\bibfnamefont {R.~Z.}\ \bibnamefont
  {Liu}}, \bibinfo {author} {\bibfnamefont {H.~S.}\ \bibnamefont {Zhong}},
  \bibinfo {author} {\bibfnamefont {Y.~M.}\ \bibnamefont {He}}, \bibinfo
  {author} {\bibfnamefont {X.}~\bibnamefont {Ding}}, \bibinfo {author}
  {\bibfnamefont {Y.~H.}\ \bibnamefont {Deng}}, \bibinfo {author}
  {\bibfnamefont {Q.}~\bibnamefont {Dai}}, \bibinfo {author} {\bibfnamefont
  {Y.~H.}\ \bibnamefont {Huo}}, \bibinfo {author} {\bibfnamefont
  {S.}~\bibnamefont {H{\"{o}}fling}}, \bibinfo {author} {\bibfnamefont {C.~Y.}\
  \bibnamefont {Lu}},\ and\ \bibinfo {author} {\bibfnamefont {J.~W.}\
  \bibnamefont {Pan}},\ }\bibfield  {title} {\bibinfo {title} {{On-Demand
  Semiconductor Source of Entangled Photons Which Simultaneously Has High
  Fidelity, Efficiency, and Indistinguishability}},\ }\href
  {https://doi.org/10.1103/PhysRevLett.122.113602} {\bibfield  {journal}
  {\bibinfo  {journal} {Phys. Rev. Lett.}\ }\textbf {\bibinfo {volume} {122}},\
  \bibinfo {pages} {113602} (\bibinfo {year} {2019})}\BibitemShut {NoStop}%
\bibitem [{\citenamefont {Schweickert}\ \emph {et~al.}(2018)\citenamefont
  {Schweickert}, \citenamefont {J{\"{o}}ns}, \citenamefont {Zeuner},
  \citenamefont {{Covre Da Silva}}, \citenamefont {Huang}, \citenamefont
  {Lettner}, \citenamefont {Reindl}, \citenamefont {Zichi}, \citenamefont
  {Trotta}, \citenamefont {Rastelli},\ and\ \citenamefont
  {Zwiller}}]{Schweickert2018}%
  \BibitemOpen
  \bibfield  {author} {\bibinfo {author} {\bibfnamefont {L.}~\bibnamefont
  {Schweickert}}, \bibinfo {author} {\bibfnamefont {K.~D.}\ \bibnamefont
  {J{\"{o}}ns}}, \bibinfo {author} {\bibfnamefont {K.~D.}\ \bibnamefont
  {Zeuner}}, \bibinfo {author} {\bibfnamefont {S.~F.}\ \bibnamefont {{Covre Da
  Silva}}}, \bibinfo {author} {\bibfnamefont {H.}~\bibnamefont {Huang}},
  \bibinfo {author} {\bibfnamefont {T.}~\bibnamefont {Lettner}}, \bibinfo
  {author} {\bibfnamefont {M.}~\bibnamefont {Reindl}}, \bibinfo {author}
  {\bibfnamefont {J.}~\bibnamefont {Zichi}}, \bibinfo {author} {\bibfnamefont
  {R.}~\bibnamefont {Trotta}}, \bibinfo {author} {\bibfnamefont
  {A.}~\bibnamefont {Rastelli}},\ and\ \bibinfo {author} {\bibfnamefont
  {V.}~\bibnamefont {Zwiller}},\ }\bibfield  {title} {\bibinfo {title}
  {{On-demand generation of background-free single photons from a solid-state
  source}},\ }\href {https://doi.org/10.1063/1.5020038} {\bibfield  {journal}
  {\bibinfo  {journal} {Appl. Phys. Lett.}\ }\textbf {\bibinfo {volume}
  {112}},\ \bibinfo {pages} {093106} (\bibinfo {year} {2018})}\BibitemShut
  {NoStop}%
\bibitem [{\citenamefont {Neuwirth}\ \emph {et~al.}(2022)\citenamefont
  {Neuwirth}, \citenamefont {Basset}, \citenamefont {Rota}, \citenamefont
  {Hartel}, \citenamefont {Sartison}, \citenamefont {da~Silva}, \citenamefont
  {J{\"{o}}ns}, \citenamefont {Rastelli},\ and\ \citenamefont
  {Trotta}}]{Neuwirth2022}%
  \BibitemOpen
  \bibfield  {author} {\bibinfo {author} {\bibfnamefont {J.}~\bibnamefont
  {Neuwirth}}, \bibinfo {author} {\bibfnamefont {F.~B.}\ \bibnamefont
  {Basset}}, \bibinfo {author} {\bibfnamefont {M.~B.}\ \bibnamefont {Rota}},
  \bibinfo {author} {\bibfnamefont {J.-G.}\ \bibnamefont {Hartel}}, \bibinfo
  {author} {\bibfnamefont {M.}~\bibnamefont {Sartison}}, \bibinfo {author}
  {\bibfnamefont {S.~F.~C.}\ \bibnamefont {da~Silva}}, \bibinfo {author}
  {\bibfnamefont {K.~D.}\ \bibnamefont {J{\"{o}}ns}}, \bibinfo {author}
  {\bibfnamefont {A.}~\bibnamefont {Rastelli}},\ and\ \bibinfo {author}
  {\bibfnamefont {R.}~\bibnamefont {Trotta}},\ }\bibfield  {title} {\bibinfo
  {title} {{A multipair-free source of entangled photons in the solid state}},\
  }\href {https://doi.org/10.1103/PhysRevB.106.L241402} {\bibfield  {journal}
  {\bibinfo  {journal} {Phys. Rev. B - Condens. Matter Mater. Phys.}\ }\textbf
  {\bibinfo {volume} {106}},\ \bibinfo {pages} {L241402} (\bibinfo {year}
  {2022})}\BibitemShut {NoStop}%
\bibitem [{\citenamefont {Vural}\ \emph {et~al.}(2020)\citenamefont {Vural},
  \citenamefont {Portalupi},\ and\ \citenamefont {Michler}}]{Vural2020}%
  \BibitemOpen
  \bibfield  {author} {\bibinfo {author} {\bibfnamefont {H.}~\bibnamefont
  {Vural}}, \bibinfo {author} {\bibfnamefont {S.~L.}\ \bibnamefont
  {Portalupi}},\ and\ \bibinfo {author} {\bibfnamefont {P.}~\bibnamefont
  {Michler}},\ }\bibfield  {title} {\bibinfo {title} {{Perspective of
  self-assembled InGaAs quantum-dots for multi-source quantum
  implementations}},\ }\href {https://doi.org/10.1063/5.0010782} {\bibfield
  {journal} {\bibinfo  {journal} {Appl. Phys. Lett.}\ }\textbf {\bibinfo
  {volume} {117}},\ \bibinfo {pages} {030501} (\bibinfo {year}
  {2020})}\BibitemShut {NoStop}%
\bibitem [{\citenamefont {Benson}\ \emph {et~al.}(2000)\citenamefont {Benson},
  \citenamefont {Santori}, \citenamefont {Pelton},\ and\ \citenamefont
  {Yamamoto}}]{Benson2000}%
  \BibitemOpen
  \bibfield  {author} {\bibinfo {author} {\bibfnamefont {O.}~\bibnamefont
  {Benson}}, \bibinfo {author} {\bibfnamefont {C.}~\bibnamefont {Santori}},
  \bibinfo {author} {\bibfnamefont {M.}~\bibnamefont {Pelton}},\ and\ \bibinfo
  {author} {\bibfnamefont {Y.}~\bibnamefont {Yamamoto}},\ }\bibfield  {title}
  {\bibinfo {title} {{Regulated and entangled photons from a single quantum
  dot}},\ }\href {https://doi.org/10.1103/PhysRevLett.84.2513} {\bibfield
  {journal} {\bibinfo  {journal} {Phys. Rev. Lett.}\ }\textbf {\bibinfo
  {volume} {84}},\ \bibinfo {pages} {2513} (\bibinfo {year}
  {2000})}\BibitemShut {NoStop}%
\bibitem [{\citenamefont {Akopian}\ \emph {et~al.}(2006)\citenamefont
  {Akopian}, \citenamefont {Lindner}, \citenamefont {Poem}, \citenamefont
  {Berlatzky}, \citenamefont {Avron}, \citenamefont {Gershoni}, \citenamefont
  {Gerardot},\ and\ \citenamefont {Petroff}}]{Akopian2006}%
  \BibitemOpen
  \bibfield  {author} {\bibinfo {author} {\bibfnamefont {N.}~\bibnamefont
  {Akopian}}, \bibinfo {author} {\bibfnamefont {N.~H.}\ \bibnamefont
  {Lindner}}, \bibinfo {author} {\bibfnamefont {E.}~\bibnamefont {Poem}},
  \bibinfo {author} {\bibfnamefont {Y.}~\bibnamefont {Berlatzky}}, \bibinfo
  {author} {\bibfnamefont {J.}~\bibnamefont {Avron}}, \bibinfo {author}
  {\bibfnamefont {D.}~\bibnamefont {Gershoni}}, \bibinfo {author}
  {\bibfnamefont {B.~D.}\ \bibnamefont {Gerardot}},\ and\ \bibinfo {author}
  {\bibfnamefont {P.~M.}\ \bibnamefont {Petroff}},\ }\bibfield  {title}
  {\bibinfo {title} {{Entangled photon pairs from semiconductor quantum
  dots}},\ }\href {https://doi.org/10.1103/PhysRevLett.96.130501} {\bibfield
  {journal} {\bibinfo  {journal} {Phys. Rev. Lett.}\ }\textbf {\bibinfo
  {volume} {96}},\ \bibinfo {pages} {130501} (\bibinfo {year}
  {2006})}\BibitemShut {NoStop}%
\bibitem [{\citenamefont {Young}\ \emph {et~al.}(2006)\citenamefont {Young},
  \citenamefont {Stevenson}, \citenamefont {Atkinson}, \citenamefont {Cooper},
  \citenamefont {Ritchie},\ and\ \citenamefont {Shields}}]{Young2006}%
  \BibitemOpen
  \bibfield  {author} {\bibinfo {author} {\bibfnamefont {R.~J.}\ \bibnamefont
  {Young}}, \bibinfo {author} {\bibfnamefont {R.~M.}\ \bibnamefont
  {Stevenson}}, \bibinfo {author} {\bibfnamefont {P.}~\bibnamefont {Atkinson}},
  \bibinfo {author} {\bibfnamefont {K.}~\bibnamefont {Cooper}}, \bibinfo
  {author} {\bibfnamefont {D.~A.}\ \bibnamefont {Ritchie}},\ and\ \bibinfo
  {author} {\bibfnamefont {A.~J.}\ \bibnamefont {Shields}},\ }\bibfield
  {title} {\bibinfo {title} {{Improved fidelity of triggered entangled photons
  from single quantum dots}},\ }\href
  {https://doi.org/10.1088/1367-2630/8/2/029} {\bibfield  {journal} {\bibinfo
  {journal} {New J. Phys.}\ }\textbf {\bibinfo {volume} {8}},\ \bibinfo {pages}
  {29} (\bibinfo {year} {2006})}\BibitemShut {NoStop}%
\bibitem [{\citenamefont {Hafenbrak}\ \emph {et~al.}(2007)\citenamefont
  {Hafenbrak}, \citenamefont {Ulrich}, \citenamefont {Michler}, \citenamefont
  {Wang}, \citenamefont {Rastelli},\ and\ \citenamefont
  {Schmidt}}]{Hafenbrak2007}%
  \BibitemOpen
  \bibfield  {author} {\bibinfo {author} {\bibfnamefont {R.}~\bibnamefont
  {Hafenbrak}}, \bibinfo {author} {\bibfnamefont {S.~M.}\ \bibnamefont
  {Ulrich}}, \bibinfo {author} {\bibfnamefont {P.}~\bibnamefont {Michler}},
  \bibinfo {author} {\bibfnamefont {L.}~\bibnamefont {Wang}}, \bibinfo {author}
  {\bibfnamefont {A.}~\bibnamefont {Rastelli}},\ and\ \bibinfo {author}
  {\bibfnamefont {O.~G.}\ \bibnamefont {Schmidt}},\ }\bibfield  {title}
  {\bibinfo {title} {{Triggered polarization-entangled photon pairs from a
  single quantum dot up to 30 K}},\ }\href
  {https://doi.org/10.1088/1367-2630/9/9/315} {\bibfield  {journal} {\bibinfo
  {journal} {New J. Phys.}\ }\textbf {\bibinfo {volume} {9}},\ \bibinfo {pages}
  {315} (\bibinfo {year} {2007})}\BibitemShut {NoStop}%
\bibitem [{\citenamefont {Salter}\ \emph {et~al.}(2010)\citenamefont {Salter},
  \citenamefont {Stevenson}, \citenamefont {Farrer}, \citenamefont {Nicoll},
  \citenamefont {Ritchie},\ and\ \citenamefont {Shields}}]{Salter2010}%
  \BibitemOpen
  \bibfield  {author} {\bibinfo {author} {\bibfnamefont {C.~L.}\ \bibnamefont
  {Salter}}, \bibinfo {author} {\bibfnamefont {R.~M.}\ \bibnamefont
  {Stevenson}}, \bibinfo {author} {\bibfnamefont {I.}~\bibnamefont {Farrer}},
  \bibinfo {author} {\bibfnamefont {C.~A.}\ \bibnamefont {Nicoll}}, \bibinfo
  {author} {\bibfnamefont {D.~A.}\ \bibnamefont {Ritchie}},\ and\ \bibinfo
  {author} {\bibfnamefont {A.~J.}\ \bibnamefont {Shields}},\ }\bibfield
  {title} {\bibinfo {title} {{An entangled-light-emitting diode}},\ }\href
  {https://doi.org/10.1038/nature09078} {\bibfield  {journal} {\bibinfo
  {journal} {Nature}\ }\textbf {\bibinfo {volume} {465}},\ \bibinfo {pages}
  {594} (\bibinfo {year} {2010})}\BibitemShut {NoStop}%
\bibitem [{\citenamefont {Dousse}\ \emph {et~al.}(2010)\citenamefont {Dousse},
  \citenamefont {Suffczy{\'{n}}ski}, \citenamefont {Beveratos}, \citenamefont
  {Krebs}, \citenamefont {Lema{\^{i}}tre}, \citenamefont {Sagnes},
  \citenamefont {Bloch}, \citenamefont {Voisin},\ and\ \citenamefont
  {Senellart}}]{Dousse2010}%
  \BibitemOpen
  \bibfield  {author} {\bibinfo {author} {\bibfnamefont {A.}~\bibnamefont
  {Dousse}}, \bibinfo {author} {\bibfnamefont {J.}~\bibnamefont
  {Suffczy{\'{n}}ski}}, \bibinfo {author} {\bibfnamefont {A.}~\bibnamefont
  {Beveratos}}, \bibinfo {author} {\bibfnamefont {O.}~\bibnamefont {Krebs}},
  \bibinfo {author} {\bibfnamefont {A.}~\bibnamefont {Lema{\^{i}}tre}},
  \bibinfo {author} {\bibfnamefont {I.}~\bibnamefont {Sagnes}}, \bibinfo
  {author} {\bibfnamefont {J.}~\bibnamefont {Bloch}}, \bibinfo {author}
  {\bibfnamefont {P.}~\bibnamefont {Voisin}},\ and\ \bibinfo {author}
  {\bibfnamefont {P.}~\bibnamefont {Senellart}},\ }\bibfield  {title} {\bibinfo
  {title} {{Ultrabright source of entangled photon pairs}},\ }\href
  {https://doi.org/10.1038/nature09148} {\bibfield  {journal} {\bibinfo
  {journal} {Nature}\ }\textbf {\bibinfo {volume} {466}},\ \bibinfo {pages}
  {217} (\bibinfo {year} {2010})}\BibitemShut {NoStop}%
\bibitem [{\citenamefont {Juska}\ \emph {et~al.}(2013)\citenamefont {Juska},
  \citenamefont {Dimastrodonato}, \citenamefont {Mereni}, \citenamefont
  {Gocalinska},\ and\ \citenamefont {Pelucchi}}]{Juska2013}%
  \BibitemOpen
  \bibfield  {author} {\bibinfo {author} {\bibfnamefont {G.}~\bibnamefont
  {Juska}}, \bibinfo {author} {\bibfnamefont {V.}~\bibnamefont
  {Dimastrodonato}}, \bibinfo {author} {\bibfnamefont {L.~O.}\ \bibnamefont
  {Mereni}}, \bibinfo {author} {\bibfnamefont {A.}~\bibnamefont {Gocalinska}},\
  and\ \bibinfo {author} {\bibfnamefont {E.}~\bibnamefont {Pelucchi}},\
  }\bibfield  {title} {\bibinfo {title} {{Towards quantum-dot arrays of
  entangled photon emitters}},\ }\href
  {https://doi.org/10.1038/nphoton.2013.128} {\bibfield  {journal} {\bibinfo
  {journal} {Nature Photonics}\ }\textbf {\bibinfo {volume} {7}},\ \bibinfo
  {pages} {527} (\bibinfo {year} {2013})}\BibitemShut {NoStop}%
\bibitem [{\citenamefont {M{\"{u}}ller}\ \emph {et~al.}(2014)\citenamefont
  {M{\"{u}}ller}, \citenamefont {Bounouar}, \citenamefont {J{\"{o}}ns},
  \citenamefont {Gl{\"{a}}ssl},\ and\ \citenamefont {Michler}}]{Muller2014}%
  \BibitemOpen
  \bibfield  {author} {\bibinfo {author} {\bibfnamefont {M.}~\bibnamefont
  {M{\"{u}}ller}}, \bibinfo {author} {\bibfnamefont {S.}~\bibnamefont
  {Bounouar}}, \bibinfo {author} {\bibfnamefont {K.~D.}\ \bibnamefont
  {J{\"{o}}ns}}, \bibinfo {author} {\bibfnamefont {M.}~\bibnamefont
  {Gl{\"{a}}ssl}},\ and\ \bibinfo {author} {\bibfnamefont {P.}~\bibnamefont
  {Michler}},\ }\bibfield  {title} {\bibinfo {title} {{On-demand generation of
  indistinguishable polarization-entangled photon pairs}},\ }\href
  {https://doi.org/10.1038/nphoton.2013.377} {\bibfield  {journal} {\bibinfo
  {journal} {Nat. Photonics}\ }\textbf {\bibinfo {volume} {8}},\ \bibinfo
  {pages} {224} (\bibinfo {year} {2014})}\BibitemShut {NoStop}%
\bibitem [{\citenamefont {Trotta}\ \emph {et~al.}(2014)\citenamefont {Trotta},
  \citenamefont {Wildmann}, \citenamefont {Zallo}, \citenamefont {Schmidt},\
  and\ \citenamefont {Rastelli}}]{Trotta2014}%
  \BibitemOpen
  \bibfield  {author} {\bibinfo {author} {\bibfnamefont {R.}~\bibnamefont
  {Trotta}}, \bibinfo {author} {\bibfnamefont {J.~S.}\ \bibnamefont
  {Wildmann}}, \bibinfo {author} {\bibfnamefont {E.}~\bibnamefont {Zallo}},
  \bibinfo {author} {\bibfnamefont {O.~G.}\ \bibnamefont {Schmidt}},\ and\
  \bibinfo {author} {\bibfnamefont {A.}~\bibnamefont {Rastelli}},\ }\bibfield
  {title} {\bibinfo {title} {{Highly entangled photons from hybrid
  piezoelectric-semiconductor quantum dot devices}},\ }\href
  {https://doi.org/10.1021/nl500968k} {\bibfield  {journal} {\bibinfo
  {journal} {Nano Lett.}\ }\textbf {\bibinfo {volume} {14}},\ \bibinfo {pages}
  {3439} (\bibinfo {year} {2014})}\BibitemShut {NoStop}%
\bibitem [{\citenamefont {Olbrich}\ \emph {et~al.}(2017)\citenamefont
  {Olbrich}, \citenamefont {H{\"{o}}schele}, \citenamefont {M{\"{u}}ller},
  \citenamefont {Kettler}, \citenamefont {{Luca Portalupi}}, \citenamefont
  {Paul}, \citenamefont {Jetter},\ and\ \citenamefont {Michler}}]{Olbrich2017}%
  \BibitemOpen
  \bibfield  {author} {\bibinfo {author} {\bibfnamefont {F.}~\bibnamefont
  {Olbrich}}, \bibinfo {author} {\bibfnamefont {J.}~\bibnamefont
  {H{\"{o}}schele}}, \bibinfo {author} {\bibfnamefont {M.}~\bibnamefont
  {M{\"{u}}ller}}, \bibinfo {author} {\bibfnamefont {J.}~\bibnamefont
  {Kettler}}, \bibinfo {author} {\bibfnamefont {S.}~\bibnamefont {{Luca
  Portalupi}}}, \bibinfo {author} {\bibfnamefont {M.}~\bibnamefont {Paul}},
  \bibinfo {author} {\bibfnamefont {M.}~\bibnamefont {Jetter}},\ and\ \bibinfo
  {author} {\bibfnamefont {P.}~\bibnamefont {Michler}},\ }\bibfield  {title}
  {\bibinfo {title} {{Polarization-entangled photons from an InGaAs-based
  quantum dot emitting in the telecom C-band}},\ }\href
  {https://doi.org/10.1063/1.4994145} {\bibfield  {journal} {\bibinfo
  {journal} {Appl. Phys. Lett.}\ }\textbf {\bibinfo {volume} {111}},\ \bibinfo
  {pages} {133106} (\bibinfo {year} {2017})}\BibitemShut {NoStop}%
\bibitem [{\citenamefont {M{\"{u}}ller}\ \emph {et~al.}(2018)\citenamefont
  {M{\"{u}}ller}, \citenamefont {Skiba-Szymanska}, \citenamefont {Krysa},
  \citenamefont {Huwer}, \citenamefont {Felle}, \citenamefont {Anderson},
  \citenamefont {Stevenson}, \citenamefont {Heffernan}, \citenamefont
  {Ritchie},\ and\ \citenamefont {Shields}}]{Mueller2018}%
  \BibitemOpen
  \bibfield  {author} {\bibinfo {author} {\bibfnamefont {T.}~\bibnamefont
  {M{\"{u}}ller}}, \bibinfo {author} {\bibfnamefont {J.}~\bibnamefont
  {Skiba-Szymanska}}, \bibinfo {author} {\bibfnamefont {A.~B.}\ \bibnamefont
  {Krysa}}, \bibinfo {author} {\bibfnamefont {J.}~\bibnamefont {Huwer}},
  \bibinfo {author} {\bibfnamefont {M.}~\bibnamefont {Felle}}, \bibinfo
  {author} {\bibfnamefont {M.}~\bibnamefont {Anderson}}, \bibinfo {author}
  {\bibfnamefont {R.~M.}\ \bibnamefont {Stevenson}}, \bibinfo {author}
  {\bibfnamefont {J.}~\bibnamefont {Heffernan}}, \bibinfo {author}
  {\bibfnamefont {D.~A.}\ \bibnamefont {Ritchie}},\ and\ \bibinfo {author}
  {\bibfnamefont {A.~J.}\ \bibnamefont {Shields}},\ }\bibfield  {title}
  {\bibinfo {title} {{A quantum light-emitting diode for the standard telecom
  window around 1,550 nm}},\ }\href
  {https://doi.org/10.1038/s41467-018-03251-7} {\bibfield  {journal} {\bibinfo
  {journal} {Nat. Commun.}\ }\textbf {\bibinfo {volume} {9}},\ \bibinfo {pages}
  {1} (\bibinfo {year} {2018})}\BibitemShut {NoStop}%
\bibitem [{\citenamefont {Zeuner}\ \emph {et~al.}(2021)\citenamefont {Zeuner},
  \citenamefont {J{\"{o}}ns}, \citenamefont {Schweickert}, \citenamefont
  {{Reuterski{\"{o}}ld Hedlund}}, \citenamefont {{Nu{\~{n}}ez Lobato}},
  \citenamefont {Lettner}, \citenamefont {Wang}, \citenamefont {Gyger},
  \citenamefont {Sch{\"{o}}ll}, \citenamefont {Steinhauer}, \citenamefont
  {Hammar},\ and\ \citenamefont {Zwiller}}]{Zeuner2021}%
  \BibitemOpen
  \bibfield  {author} {\bibinfo {author} {\bibfnamefont {K.~D.}\ \bibnamefont
  {Zeuner}}, \bibinfo {author} {\bibfnamefont {K.~D.}\ \bibnamefont
  {J{\"{o}}ns}}, \bibinfo {author} {\bibfnamefont {L.}~\bibnamefont
  {Schweickert}}, \bibinfo {author} {\bibfnamefont {C.}~\bibnamefont
  {{Reuterski{\"{o}}ld Hedlund}}}, \bibinfo {author} {\bibfnamefont
  {C.}~\bibnamefont {{Nu{\~{n}}ez Lobato}}}, \bibinfo {author} {\bibfnamefont
  {T.}~\bibnamefont {Lettner}}, \bibinfo {author} {\bibfnamefont
  {K.}~\bibnamefont {Wang}}, \bibinfo {author} {\bibfnamefont {S.}~\bibnamefont
  {Gyger}}, \bibinfo {author} {\bibfnamefont {E.}~\bibnamefont {Sch{\"{o}}ll}},
  \bibinfo {author} {\bibfnamefont {S.}~\bibnamefont {Steinhauer}}, \bibinfo
  {author} {\bibfnamefont {M.}~\bibnamefont {Hammar}},\ and\ \bibinfo {author}
  {\bibfnamefont {V.}~\bibnamefont {Zwiller}},\ }\bibfield  {title} {\bibinfo
  {title} {{On-Demand Generation of Entangled Photon Pairs in the Telecom
  C-Band with InAs Quantum Dots}},\ }\href
  {https://doi.org/10.1021/acsphotonics.1c00504} {\bibfield  {journal}
  {\bibinfo  {journal} {ACS Photonics}\ }\textbf {\bibinfo {volume} {8}},\
  \bibinfo {pages} {2337} (\bibinfo {year} {2021})}\BibitemShut {NoStop}%
\bibitem [{\citenamefont {Winik}\ \emph {et~al.}(2017)\citenamefont {Winik},
  \citenamefont {Cogan}, \citenamefont {Don}, \citenamefont {Schwartz},
  \citenamefont {Gantz}, \citenamefont {Schmidgall}, \citenamefont {Livneh},
  \citenamefont {Rapaport}, \citenamefont {Buks},\ and\ \citenamefont
  {Gershoni}}]{Winik2017}%
  \BibitemOpen
  \bibfield  {author} {\bibinfo {author} {\bibfnamefont {R.}~\bibnamefont
  {Winik}}, \bibinfo {author} {\bibfnamefont {D.}~\bibnamefont {Cogan}},
  \bibinfo {author} {\bibfnamefont {Y.}~\bibnamefont {Don}}, \bibinfo {author}
  {\bibfnamefont {I.}~\bibnamefont {Schwartz}}, \bibinfo {author}
  {\bibfnamefont {L.}~\bibnamefont {Gantz}}, \bibinfo {author} {\bibfnamefont
  {E.~R.}\ \bibnamefont {Schmidgall}}, \bibinfo {author} {\bibfnamefont
  {N.}~\bibnamefont {Livneh}}, \bibinfo {author} {\bibfnamefont
  {R.}~\bibnamefont {Rapaport}}, \bibinfo {author} {\bibfnamefont
  {E.}~\bibnamefont {Buks}},\ and\ \bibinfo {author} {\bibfnamefont
  {D.}~\bibnamefont {Gershoni}},\ }\bibfield  {title} {\bibinfo {title}
  {{On-demand source of maximally entangled photon pairs using the
  biexciton-exciton radiative cascade}},\ }\href
  {https://doi.org/10.1103/PhysRevB.95.235435} {\bibfield  {journal} {\bibinfo
  {journal} {Phys. Rev. B}\ }\textbf {\bibinfo {volume} {95}},\ \bibinfo
  {pages} {235435} (\bibinfo {year} {2017})}\BibitemShut {NoStop}%
\bibitem [{\citenamefont {Fognini}\ \emph {et~al.}(2019)\citenamefont
  {Fognini}, \citenamefont {Ahmadi}, \citenamefont {Zeeshan}, \citenamefont
  {Fokkens}, \citenamefont {Gibson}, \citenamefont {Sherlekar}, \citenamefont
  {Daley}, \citenamefont {Dalacu}, \citenamefont {Poole}, \citenamefont
  {J{\"{o}}ns}, \citenamefont {Zwiller},\ and\ \citenamefont
  {Reimer}}]{Fognini2019}%
  \BibitemOpen
  \bibfield  {author} {\bibinfo {author} {\bibfnamefont {A.}~\bibnamefont
  {Fognini}}, \bibinfo {author} {\bibfnamefont {A.}~\bibnamefont {Ahmadi}},
  \bibinfo {author} {\bibfnamefont {M.}~\bibnamefont {Zeeshan}}, \bibinfo
  {author} {\bibfnamefont {J.~T.}\ \bibnamefont {Fokkens}}, \bibinfo {author}
  {\bibfnamefont {S.~J.}\ \bibnamefont {Gibson}}, \bibinfo {author}
  {\bibfnamefont {N.}~\bibnamefont {Sherlekar}}, \bibinfo {author}
  {\bibfnamefont {S.~J.}\ \bibnamefont {Daley}}, \bibinfo {author}
  {\bibfnamefont {D.}~\bibnamefont {Dalacu}}, \bibinfo {author} {\bibfnamefont
  {P.~J.}\ \bibnamefont {Poole}}, \bibinfo {author} {\bibfnamefont {K.~D.}\
  \bibnamefont {J{\"{o}}ns}}, \bibinfo {author} {\bibfnamefont
  {V.}~\bibnamefont {Zwiller}},\ and\ \bibinfo {author} {\bibfnamefont {M.~E.}\
  \bibnamefont {Reimer}},\ }\bibfield  {title} {\bibinfo {title} {{Dephasing
  Free Photon Entanglement with a Quantum Dot}},\ }\href
  {https://doi.org/10.1021/acsphotonics.8b01496} {\bibfield  {journal}
  {\bibinfo  {journal} {ACS Photonics}\ }\textbf {\bibinfo {volume} {6}},\
  \bibinfo {pages} {1656} (\bibinfo {year} {2019})}\BibitemShut {NoStop}%
\bibitem [{\citenamefont {Anderson}\ \emph {et~al.}(2020)\citenamefont
  {Anderson}, \citenamefont {M{\"{u}}ller}, \citenamefont {Skiba-Szymanska},
  \citenamefont {Krysa}, \citenamefont {Huwer}, \citenamefont {Stevenson},
  \citenamefont {Heffernan}, \citenamefont {Ritchie},\ and\ \citenamefont
  {Shields}}]{Anderson2020}%
  \BibitemOpen
  \bibfield  {author} {\bibinfo {author} {\bibfnamefont {M.}~\bibnamefont
  {Anderson}}, \bibinfo {author} {\bibfnamefont {T.}~\bibnamefont
  {M{\"{u}}ller}}, \bibinfo {author} {\bibfnamefont {J.}~\bibnamefont
  {Skiba-Szymanska}}, \bibinfo {author} {\bibfnamefont {A.~B.}\ \bibnamefont
  {Krysa}}, \bibinfo {author} {\bibfnamefont {J.}~\bibnamefont {Huwer}},
  \bibinfo {author} {\bibfnamefont {R.~M.}\ \bibnamefont {Stevenson}}, \bibinfo
  {author} {\bibfnamefont {J.}~\bibnamefont {Heffernan}}, \bibinfo {author}
  {\bibfnamefont {D.~A.}\ \bibnamefont {Ritchie}},\ and\ \bibinfo {author}
  {\bibfnamefont {A.~J.}\ \bibnamefont {Shields}},\ }\bibfield  {title}
  {\bibinfo {title} {{Gigahertz-Clocked Teleportation of Time-Bin Qubits with a
  Quantum Dot in the Telecommunication C Band}},\ }\href
  {https://doi.org/10.1103/PhysRevApplied.13.054052} {\bibfield  {journal}
  {\bibinfo  {journal} {Phys. Rev. Appl.}\ }\textbf {\bibinfo {volume} {13}},\
  \bibinfo {pages} {054052} (\bibinfo {year} {2020})}\BibitemShut {NoStop}%
\bibitem [{\citenamefont {Huber}\ \emph {et~al.}(2018)\citenamefont {Huber},
  \citenamefont {Reindl}, \citenamefont {{Covre da Silva}}, \citenamefont
  {Schimpf}, \citenamefont {Mart{\'{i}}n-S{\'{a}}nchez}, \citenamefont {Huang},
  \citenamefont {Piredda}, \citenamefont {Edlinger}, \citenamefont {Rastelli},\
  and\ \citenamefont {Trotta}}]{Huber2018}%
  \BibitemOpen
  \bibfield  {author} {\bibinfo {author} {\bibfnamefont {D.}~\bibnamefont
  {Huber}}, \bibinfo {author} {\bibfnamefont {M.}~\bibnamefont {Reindl}},
  \bibinfo {author} {\bibfnamefont {S.~F.}\ \bibnamefont {{Covre da Silva}}},
  \bibinfo {author} {\bibfnamefont {C.}~\bibnamefont {Schimpf}}, \bibinfo
  {author} {\bibfnamefont {J.}~\bibnamefont {Mart{\'{i}}n-S{\'{a}}nchez}},
  \bibinfo {author} {\bibfnamefont {H.}~\bibnamefont {Huang}}, \bibinfo
  {author} {\bibfnamefont {G.}~\bibnamefont {Piredda}}, \bibinfo {author}
  {\bibfnamefont {J.}~\bibnamefont {Edlinger}}, \bibinfo {author}
  {\bibfnamefont {A.}~\bibnamefont {Rastelli}},\ and\ \bibinfo {author}
  {\bibfnamefont {R.}~\bibnamefont {Trotta}},\ }\bibfield  {title} {\bibinfo
  {title} {{Strain-Tunable GaAs Quantum Dot: A Nearly Dephasing-Free Source of
  Entangled Photon Pairs on Demand}},\ }\href
  {https://doi.org/10.1103/PhysRevLett.121.033902} {\bibfield  {journal}
  {\bibinfo  {journal} {Phys. Rev. Lett.}\ }\textbf {\bibinfo {volume} {121}},\
  \bibinfo {pages} {033902} (\bibinfo {year} {2018})}\BibitemShut {NoStop}%
\bibitem [{\citenamefont {Gurioli}\ \emph {et~al.}(2019)\citenamefont
  {Gurioli}, \citenamefont {Wang}, \citenamefont {Rastelli}, \citenamefont
  {Kuroda},\ and\ \citenamefont {Sanguinetti}}]{Gurioli2019}%
  \BibitemOpen
  \bibfield  {author} {\bibinfo {author} {\bibfnamefont {M.}~\bibnamefont
  {Gurioli}}, \bibinfo {author} {\bibfnamefont {Z.}~\bibnamefont {Wang}},
  \bibinfo {author} {\bibfnamefont {A.}~\bibnamefont {Rastelli}}, \bibinfo
  {author} {\bibfnamefont {T.}~\bibnamefont {Kuroda}},\ and\ \bibinfo {author}
  {\bibfnamefont {S.}~\bibnamefont {Sanguinetti}},\ }\bibfield  {title}
  {\bibinfo {title} {{Droplet epitaxy of semiconductor nanostructures for
  quantum photonic devices}},\ }\href
  {https://doi.org/10.1038/s41563-019-0355-y} {\bibfield  {journal} {\bibinfo
  {journal} {Nat. Mater.}\ }\textbf {\bibinfo {volume} {18}},\ \bibinfo {pages}
  {799} (\bibinfo {year} {2019})}\BibitemShut {NoStop}%
\bibitem [{\citenamefont {da~Silva}\ \emph {et~al.}(2021)\citenamefont
  {da~Silva}, \citenamefont {Undeutsch}, \citenamefont {Lehner}, \citenamefont
  {Manna}, \citenamefont {Krieger}, \citenamefont {Reindl}, \citenamefont
  {Schimpf}, \citenamefont {Trotta},\ and\ \citenamefont
  {Rastelli}}]{DaSilva2021}%
  \BibitemOpen
  \bibfield  {author} {\bibinfo {author} {\bibfnamefont {S.~F.~C.}\
  \bibnamefont {da~Silva}}, \bibinfo {author} {\bibfnamefont {G.}~\bibnamefont
  {Undeutsch}}, \bibinfo {author} {\bibfnamefont {B.}~\bibnamefont {Lehner}},
  \bibinfo {author} {\bibfnamefont {S.}~\bibnamefont {Manna}}, \bibinfo
  {author} {\bibfnamefont {T.~M.}\ \bibnamefont {Krieger}}, \bibinfo {author}
  {\bibfnamefont {M.}~\bibnamefont {Reindl}}, \bibinfo {author} {\bibfnamefont
  {C.}~\bibnamefont {Schimpf}}, \bibinfo {author} {\bibfnamefont
  {R.}~\bibnamefont {Trotta}},\ and\ \bibinfo {author} {\bibfnamefont
  {A.}~\bibnamefont {Rastelli}},\ }\bibfield  {title} {\bibinfo {title} {{GaAs
  quantum dots grown by droplet etching epitaxy as quantum light sources}},\
  }\href {https://doi.org/10.1063/5.0057070} {\bibfield  {journal} {\bibinfo
  {journal} {Appl. Phys. Lett.}\ }\textbf {\bibinfo {volume} {119}},\ \bibinfo
  {pages} {120502} (\bibinfo {year} {2021})}\BibitemShut {NoStop}%
\bibitem [{\citenamefont {Zhai}\ \emph {et~al.}(2022)\citenamefont {Zhai},
  \citenamefont {Nguyen}, \citenamefont {Spinnler}, \citenamefont {Ritzmann},
  \citenamefont {L{\"{o}}bl}, \citenamefont {Wieck}, \citenamefont {Ludwig},
  \citenamefont {Javadi},\ and\ \citenamefont {Warburton}}]{Zhai2022}%
  \BibitemOpen
  \bibfield  {author} {\bibinfo {author} {\bibfnamefont {L.}~\bibnamefont
  {Zhai}}, \bibinfo {author} {\bibfnamefont {G.~N.}\ \bibnamefont {Nguyen}},
  \bibinfo {author} {\bibfnamefont {C.}~\bibnamefont {Spinnler}}, \bibinfo
  {author} {\bibfnamefont {J.}~\bibnamefont {Ritzmann}}, \bibinfo {author}
  {\bibfnamefont {M.~C.}\ \bibnamefont {L{\"{o}}bl}}, \bibinfo {author}
  {\bibfnamefont {A.~D.}\ \bibnamefont {Wieck}}, \bibinfo {author}
  {\bibfnamefont {A.}~\bibnamefont {Ludwig}}, \bibinfo {author} {\bibfnamefont
  {A.}~\bibnamefont {Javadi}},\ and\ \bibinfo {author} {\bibfnamefont {R.~J.}\
  \bibnamefont {Warburton}},\ }\bibfield  {title} {\bibinfo {title} {{Quantum
  interference of identical photons from remote GaAs quantum dots}},\ }\href
  {https://doi.org/10.1038/s41565-022-01131-2} {\bibfield  {journal} {\bibinfo
  {journal} {Nature Nanotechnology}\ ,\ \bibinfo {pages} {1}} (\bibinfo {year}
  {2022})},\ \Eprint {https://arxiv.org/abs/2106.03871} {arXiv:2106.03871}
  \BibitemShut {NoStop}%
\bibitem [{\citenamefont {Bayer}\ \emph {et~al.}(2002)\citenamefont {Bayer},
  \citenamefont {Ortner}, \citenamefont {Stern}, \citenamefont {Kuther},
  \citenamefont {Gorbunov}, \citenamefont {Forchel}, \citenamefont {Hawrylak},
  \citenamefont {Fafard}, \citenamefont {Hinzer}, \citenamefont {Reinecke},
  \citenamefont {Walck}, \citenamefont {Reithmaier}, \citenamefont {Klopf},\
  and\ \citenamefont {Sch{\"{a}}fer}}]{Bayer2002}%
  \BibitemOpen
  \bibfield  {author} {\bibinfo {author} {\bibfnamefont {M.}~\bibnamefont
  {Bayer}}, \bibinfo {author} {\bibfnamefont {G.}~\bibnamefont {Ortner}},
  \bibinfo {author} {\bibfnamefont {O.}~\bibnamefont {Stern}}, \bibinfo
  {author} {\bibfnamefont {A.}~\bibnamefont {Kuther}}, \bibinfo {author}
  {\bibfnamefont {A.~A.}\ \bibnamefont {Gorbunov}}, \bibinfo {author}
  {\bibfnamefont {A.}~\bibnamefont {Forchel}}, \bibinfo {author} {\bibfnamefont
  {P.}~\bibnamefont {Hawrylak}}, \bibinfo {author} {\bibfnamefont
  {S.}~\bibnamefont {Fafard}}, \bibinfo {author} {\bibfnamefont
  {K.}~\bibnamefont {Hinzer}}, \bibinfo {author} {\bibfnamefont {T.~L.}\
  \bibnamefont {Reinecke}}, \bibinfo {author} {\bibfnamefont {S.~N.}\
  \bibnamefont {Walck}}, \bibinfo {author} {\bibfnamefont {J.~P.}\ \bibnamefont
  {Reithmaier}}, \bibinfo {author} {\bibfnamefont {F.}~\bibnamefont {Klopf}},\
  and\ \bibinfo {author} {\bibfnamefont {F.}~\bibnamefont {Sch{\"{a}}fer}},\
  }\bibfield  {title} {\bibinfo {title} {{Fine structure of neutral and charged
  excitons in self-assembled In(Ga)As/(Al)GaAs quantum dots}},\ }\href
  {https://doi.org/10.1103/PhysRevB.65.195315} {\bibfield  {journal} {\bibinfo
  {journal} {Phys. Rev. B}\ }\textbf {\bibinfo {volume} {65}},\ \bibinfo
  {pages} {195315} (\bibinfo {year} {2002})}\BibitemShut {NoStop}%
\bibitem [{\citenamefont {Hudson}\ \emph {et~al.}(2007)\citenamefont {Hudson},
  \citenamefont {Stevenson}, \citenamefont {Bennett}, \citenamefont {Young},
  \citenamefont {Nicoll}, \citenamefont {Atkinson}, \citenamefont {Cooper},
  \citenamefont {Ritchie},\ and\ \citenamefont {Shields}}]{Hudson2007}%
  \BibitemOpen
  \bibfield  {author} {\bibinfo {author} {\bibfnamefont {A.~J.}\ \bibnamefont
  {Hudson}}, \bibinfo {author} {\bibfnamefont {R.~M.}\ \bibnamefont
  {Stevenson}}, \bibinfo {author} {\bibfnamefont {A.~J.}\ \bibnamefont
  {Bennett}}, \bibinfo {author} {\bibfnamefont {R.~J.}\ \bibnamefont {Young}},
  \bibinfo {author} {\bibfnamefont {C.~A.}\ \bibnamefont {Nicoll}}, \bibinfo
  {author} {\bibfnamefont {P.}~\bibnamefont {Atkinson}}, \bibinfo {author}
  {\bibfnamefont {K.}~\bibnamefont {Cooper}}, \bibinfo {author} {\bibfnamefont
  {D.~A.}\ \bibnamefont {Ritchie}},\ and\ \bibinfo {author} {\bibfnamefont
  {A.~J.}\ \bibnamefont {Shields}},\ }\bibfield  {title} {\bibinfo {title}
  {Coherence of an entangled exciton-photon state},\ }\href
  {https://doi.org/10.1103/PhysRevLett.99.266802} {\bibfield  {journal}
  {\bibinfo  {journal} {Phys. Rev. Lett.}\ }\textbf {\bibinfo {volume} {99}},\
  \bibinfo {pages} {266802} (\bibinfo {year} {2007})}\BibitemShut {NoStop}%
\bibitem [{\citenamefont {Hohenester}\ \emph {et~al.}(2007)\citenamefont
  {Hohenester}, \citenamefont {Pfanner},\ and\ \citenamefont
  {Seliger}}]{Hohenester2007}%
  \BibitemOpen
  \bibfield  {author} {\bibinfo {author} {\bibfnamefont {U.}~\bibnamefont
  {Hohenester}}, \bibinfo {author} {\bibfnamefont {G.}~\bibnamefont
  {Pfanner}},\ and\ \bibinfo {author} {\bibfnamefont {M.}~\bibnamefont
  {Seliger}},\ }\bibfield  {title} {\bibinfo {title} {{Phonon-assisted
  decoherence in the production of polarization-entangled photons in a single
  semiconductor quantum dot}},\ }\href
  {https://doi.org/10.1103/PhysRevLett.99.047402} {\bibfield  {journal}
  {\bibinfo  {journal} {Phys. Rev. Lett.}\ }\textbf {\bibinfo {volume} {99}},\
  \bibinfo {pages} {047402} (\bibinfo {year} {2007})}\BibitemShut {NoStop}%
\bibitem [{\citenamefont {Reigue}\ \emph {et~al.}(2017)\citenamefont {Reigue},
  \citenamefont {Iles-Smith}, \citenamefont {Lux}, \citenamefont {Monniello},
  \citenamefont {Bernard}, \citenamefont {Margaillan}, \citenamefont
  {Lemaitre}, \citenamefont {Martinez}, \citenamefont {McCutcheon},
  \citenamefont {M{\o}rk}, \citenamefont {Hostein},\ and\ \citenamefont
  {Voliotis}}]{Reigue2017}%
  \BibitemOpen
  \bibfield  {author} {\bibinfo {author} {\bibfnamefont {A.}~\bibnamefont
  {Reigue}}, \bibinfo {author} {\bibfnamefont {J.}~\bibnamefont {Iles-Smith}},
  \bibinfo {author} {\bibfnamefont {F.}~\bibnamefont {Lux}}, \bibinfo {author}
  {\bibfnamefont {L.}~\bibnamefont {Monniello}}, \bibinfo {author}
  {\bibfnamefont {M.}~\bibnamefont {Bernard}}, \bibinfo {author} {\bibfnamefont
  {F.}~\bibnamefont {Margaillan}}, \bibinfo {author} {\bibfnamefont
  {A.}~\bibnamefont {Lemaitre}}, \bibinfo {author} {\bibfnamefont
  {A.}~\bibnamefont {Martinez}}, \bibinfo {author} {\bibfnamefont {D.~P.}\
  \bibnamefont {McCutcheon}}, \bibinfo {author} {\bibfnamefont
  {J.}~\bibnamefont {M{\o}rk}}, \bibinfo {author} {\bibfnamefont
  {R.}~\bibnamefont {Hostein}},\ and\ \bibinfo {author} {\bibfnamefont
  {V.}~\bibnamefont {Voliotis}},\ }\bibfield  {title} {\bibinfo {title}
  {{Probing Electron-Phonon Interaction through Two-Photon Interference in
  Resonantly Driven Semiconductor Quantum Dots}},\ }\href
  {https://doi.org/10.1103/PhysRevLett.118.233602} {\bibfield  {journal}
  {\bibinfo  {journal} {Phys. Rev. Lett.}\ }\textbf {\bibinfo {volume} {118}},\
  \bibinfo {pages} {233602} (\bibinfo {year} {2017})}\BibitemShut {NoStop}%
\bibitem [{\citenamefont {Seidelmann}\ \emph {et~al.}(2022)\citenamefont
  {Seidelmann}, \citenamefont {Schimpf}, \citenamefont {Bracht}, \citenamefont
  {Cosacchi}, \citenamefont {Vagov}, \citenamefont {Rastelli}, \citenamefont
  {Reiter},\ and\ \citenamefont {Axt}}]{Seidelmann2022}%
  \BibitemOpen
  \bibfield  {author} {\bibinfo {author} {\bibfnamefont {T.}~\bibnamefont
  {Seidelmann}}, \bibinfo {author} {\bibfnamefont {C.}~\bibnamefont {Schimpf}},
  \bibinfo {author} {\bibfnamefont {T.~K.}\ \bibnamefont {Bracht}}, \bibinfo
  {author} {\bibfnamefont {M.}~\bibnamefont {Cosacchi}}, \bibinfo {author}
  {\bibfnamefont {A.}~\bibnamefont {Vagov}}, \bibinfo {author} {\bibfnamefont
  {A.}~\bibnamefont {Rastelli}}, \bibinfo {author} {\bibfnamefont {D.~E.}\
  \bibnamefont {Reiter}},\ and\ \bibinfo {author} {\bibfnamefont {V.~M.}\
  \bibnamefont {Axt}},\ }\bibfield  {title} {\bibinfo {title} {{Two-Photon
  Excitation Sets Fundamental Limit to Entangled Photon Pair Generation from
  Quantum Emitters}},\ }\href {https://doi.org/10.1103/PhysRevLett.129.193604}
  {\bibfield  {journal} {\bibinfo  {journal} {Phys. Rev. Lett.}\ }\textbf
  {\bibinfo {volume} {129}},\ \bibinfo {pages} {193604} (\bibinfo {year}
  {2022})}\BibitemShut {NoStop}%
\bibitem [{\citenamefont {Seidelmann}\ \emph {et~al.}(2023)\citenamefont
  {Seidelmann}, \citenamefont {Bracht}, \citenamefont {Lehner}, \citenamefont
  {Schimpf}, \citenamefont {Cosacchi}, \citenamefont {Cygorek}, \citenamefont
  {Vagov}, \citenamefont {Rastelli}, \citenamefont {Reiter},\ and\
  \citenamefont {Axt}}]{Seidelmann2023}%
  \BibitemOpen
  \bibfield  {author} {\bibinfo {author} {\bibfnamefont {T.}~\bibnamefont
  {Seidelmann}}, \bibinfo {author} {\bibfnamefont {T.~K.}\ \bibnamefont
  {Bracht}}, \bibinfo {author} {\bibfnamefont {B.~U.}\ \bibnamefont {Lehner}},
  \bibinfo {author} {\bibfnamefont {C.}~\bibnamefont {Schimpf}}, \bibinfo
  {author} {\bibfnamefont {M.}~\bibnamefont {Cosacchi}}, \bibinfo {author}
  {\bibfnamefont {M.}~\bibnamefont {Cygorek}}, \bibinfo {author} {\bibfnamefont
  {A.}~\bibnamefont {Vagov}}, \bibinfo {author} {\bibfnamefont
  {A.}~\bibnamefont {Rastelli}}, \bibinfo {author} {\bibfnamefont {D.~E.}\
  \bibnamefont {Reiter}},\ and\ \bibinfo {author} {\bibfnamefont {V.~M.}\
  \bibnamefont {Axt}},\ }\href {https://doi.org/10.48550/ARXIV.2301.10820}
  {\bibinfo {title} {Two-photon excitation with finite pulses unlocks pure
  dephasing-induced degradation of entangled photons emitted by quantum dots}}
  (\bibinfo {year} {2023}),\ \Eprint {https://arxiv.org/abs/arXiv.2301.10820}
  {arXiv.2301.10820} \BibitemShut {NoStop}%
\bibitem [{\citenamefont {Merkulov}\ \emph {et~al.}(2002)\citenamefont
  {Merkulov}, \citenamefont {Efros},\ and\ \citenamefont
  {Rosen}}]{Merkulov2002}%
  \BibitemOpen
  \bibfield  {author} {\bibinfo {author} {\bibfnamefont {I.~A.}\ \bibnamefont
  {Merkulov}}, \bibinfo {author} {\bibfnamefont {A.~L.}\ \bibnamefont
  {Efros}},\ and\ \bibinfo {author} {\bibfnamefont {M.}~\bibnamefont {Rosen}},\
  }\bibfield  {title} {\bibinfo {title} {{Electron spin relaxation by nuclei in
  semiconductor quantum dots}},\ }\href
  {https://doi.org/10.1103/PhysRevB.65.205309} {\bibfield  {journal} {\bibinfo
  {journal} {Phys. Rev. B - Condens. Matter Mater. Phys.}\ }\textbf {\bibinfo
  {volume} {65}},\ \bibinfo {pages} {205309} (\bibinfo {year}
  {2002})}\BibitemShut {NoStop}%
\bibitem [{\citenamefont {Urbaszek}\ \emph {et~al.}(2013)\citenamefont
  {Urbaszek}, \citenamefont {Marie}, \citenamefont {Amand}, \citenamefont
  {Krebs}, \citenamefont {Voisin}, \citenamefont {Maletinsky}, \citenamefont
  {H{\"{o}}gele},\ and\ \citenamefont {Imamoglu}}]{Urbaszek2013}%
  \BibitemOpen
  \bibfield  {author} {\bibinfo {author} {\bibfnamefont {B.}~\bibnamefont
  {Urbaszek}}, \bibinfo {author} {\bibfnamefont {X.}~\bibnamefont {Marie}},
  \bibinfo {author} {\bibfnamefont {T.}~\bibnamefont {Amand}}, \bibinfo
  {author} {\bibfnamefont {O.}~\bibnamefont {Krebs}}, \bibinfo {author}
  {\bibfnamefont {P.}~\bibnamefont {Voisin}}, \bibinfo {author} {\bibfnamefont
  {P.}~\bibnamefont {Maletinsky}}, \bibinfo {author} {\bibfnamefont
  {A.}~\bibnamefont {H{\"{o}}gele}},\ and\ \bibinfo {author} {\bibfnamefont
  {A.}~\bibnamefont {Imamoglu}},\ }\bibfield  {title} {\bibinfo {title}
  {{Nuclear spin physics in quantum dots: An optical investigation}},\ }\href
  {https://doi.org/10.1103/RevModPhys.85.79} {\bibfield  {journal} {\bibinfo
  {journal} {Rev. Mod. Phys.}\ }\textbf {\bibinfo {volume} {85}},\ \bibinfo
  {pages} {79} (\bibinfo {year} {2013})}\BibitemShut {NoStop}%
\bibitem [{\citenamefont {Kuhlmann}\ \emph {et~al.}(2013)\citenamefont
  {Kuhlmann}, \citenamefont {Houel}, \citenamefont {Ludwig}, \citenamefont
  {Greuter}, \citenamefont {Reuter}, \citenamefont {Wieck}, \citenamefont
  {Poggio},\ and\ \citenamefont {Warburton}}]{Kuhlmann2013}%
  \BibitemOpen
  \bibfield  {author} {\bibinfo {author} {\bibfnamefont {A.~V.}\ \bibnamefont
  {Kuhlmann}}, \bibinfo {author} {\bibfnamefont {J.}~\bibnamefont {Houel}},
  \bibinfo {author} {\bibfnamefont {A.}~\bibnamefont {Ludwig}}, \bibinfo
  {author} {\bibfnamefont {L.}~\bibnamefont {Greuter}}, \bibinfo {author}
  {\bibfnamefont {D.}~\bibnamefont {Reuter}}, \bibinfo {author} {\bibfnamefont
  {A.~D.}\ \bibnamefont {Wieck}}, \bibinfo {author} {\bibfnamefont
  {M.}~\bibnamefont {Poggio}},\ and\ \bibinfo {author} {\bibfnamefont {R.~J.}\
  \bibnamefont {Warburton}},\ }\bibfield  {title} {\bibinfo {title} {{Charge
  noise and spin noise in a semiconductor quantum device}},\ }\href
  {https://doi.org/10.1038/nphys2688} {\bibfield  {journal} {\bibinfo
  {journal} {Nat. Phys.}\ }\textbf {\bibinfo {volume} {9}},\ \bibinfo {pages}
  {570} (\bibinfo {year} {2013})}\BibitemShut {NoStop}%
\bibitem [{\citenamefont {Kuroda}\ \emph {et~al.}(2013)\citenamefont {Kuroda},
  \citenamefont {Mano}, \citenamefont {Ha}, \citenamefont {Nakajima},
  \citenamefont {Kumano}, \citenamefont {Urbaszek}, \citenamefont {Jo},
  \citenamefont {Abbarchi}, \citenamefont {Sakuma}, \citenamefont {Sakoda},
  \citenamefont {Suemune}, \citenamefont {Marie},\ and\ \citenamefont
  {Amand}}]{Kuroda2013}%
  \BibitemOpen
  \bibfield  {author} {\bibinfo {author} {\bibfnamefont {T.}~\bibnamefont
  {Kuroda}}, \bibinfo {author} {\bibfnamefont {T.}~\bibnamefont {Mano}},
  \bibinfo {author} {\bibfnamefont {N.}~\bibnamefont {Ha}}, \bibinfo {author}
  {\bibfnamefont {H.}~\bibnamefont {Nakajima}}, \bibinfo {author}
  {\bibfnamefont {H.}~\bibnamefont {Kumano}}, \bibinfo {author} {\bibfnamefont
  {B.}~\bibnamefont {Urbaszek}}, \bibinfo {author} {\bibfnamefont
  {M.}~\bibnamefont {Jo}}, \bibinfo {author} {\bibfnamefont {M.}~\bibnamefont
  {Abbarchi}}, \bibinfo {author} {\bibfnamefont {Y.}~\bibnamefont {Sakuma}},
  \bibinfo {author} {\bibfnamefont {K.}~\bibnamefont {Sakoda}}, \bibinfo
  {author} {\bibfnamefont {I.}~\bibnamefont {Suemune}}, \bibinfo {author}
  {\bibfnamefont {X.}~\bibnamefont {Marie}},\ and\ \bibinfo {author}
  {\bibfnamefont {T.}~\bibnamefont {Amand}},\ }\bibfield  {title} {\bibinfo
  {title} {{Symmetric quantum dots as efficient sources of highly entangled
  photons: Violation of Bell's inequality without spectral and temporal
  filtering}},\ }\href {https://doi.org/10.1103/PhysRevB.88.041306} {\bibfield
  {journal} {\bibinfo  {journal} {Physical Review B}\ }\textbf {\bibinfo
  {volume} {88}},\ \bibinfo {pages} {041306} (\bibinfo {year}
  {2013})}\BibitemShut {NoStop}%
\bibitem [{\citenamefont {Huber}\ \emph {et~al.}(2017)\citenamefont {Huber},
  \citenamefont {Reindl}, \citenamefont {Huo}, \citenamefont {Huang},
  \citenamefont {Wildmann}, \citenamefont {Schmidt}, \citenamefont {Rastelli},\
  and\ \citenamefont {Trotta}}]{Huber2017}%
  \BibitemOpen
  \bibfield  {author} {\bibinfo {author} {\bibfnamefont {D.}~\bibnamefont
  {Huber}}, \bibinfo {author} {\bibfnamefont {M.}~\bibnamefont {Reindl}},
  \bibinfo {author} {\bibfnamefont {Y.}~\bibnamefont {Huo}}, \bibinfo {author}
  {\bibfnamefont {H.}~\bibnamefont {Huang}}, \bibinfo {author} {\bibfnamefont
  {J.~S.}\ \bibnamefont {Wildmann}}, \bibinfo {author} {\bibfnamefont {O.~G.}\
  \bibnamefont {Schmidt}}, \bibinfo {author} {\bibfnamefont {A.}~\bibnamefont
  {Rastelli}},\ and\ \bibinfo {author} {\bibfnamefont {R.}~\bibnamefont
  {Trotta}},\ }\bibfield  {title} {\bibinfo {title} {{Highly indistinguishable
  and strongly entangled photons from symmetric GaAs quantum dots}},\ }\href
  {https://doi.org/10.1038/ncomms8662} {\bibfield  {journal} {\bibinfo
  {journal} {Nat. Commun.}\ }\textbf {\bibinfo {volume} {8}},\ \bibinfo {pages}
  {15506} (\bibinfo {year} {2017})}\BibitemShut {NoStop}%
\bibitem [{\citenamefont {Keil}\ \emph {et~al.}(2017)\citenamefont {Keil},
  \citenamefont {Zopf}, \citenamefont {Chen}, \citenamefont {H{\"{o}}fer},
  \citenamefont {Zhang}, \citenamefont {Ding},\ and\ \citenamefont
  {Schmidt}}]{keil2017}%
  \BibitemOpen
  \bibfield  {author} {\bibinfo {author} {\bibfnamefont {R.}~\bibnamefont
  {Keil}}, \bibinfo {author} {\bibfnamefont {M.}~\bibnamefont {Zopf}}, \bibinfo
  {author} {\bibfnamefont {Y.}~\bibnamefont {Chen}}, \bibinfo {author}
  {\bibfnamefont {B.}~\bibnamefont {H{\"{o}}fer}}, \bibinfo {author}
  {\bibfnamefont {J.}~\bibnamefont {Zhang}}, \bibinfo {author} {\bibfnamefont
  {F.}~\bibnamefont {Ding}},\ and\ \bibinfo {author} {\bibfnamefont {O.~G.}\
  \bibnamefont {Schmidt}},\ }\bibfield  {title} {\bibinfo {title} {{Solid-state
  ensemble of highly entangled photon sources at rubidium atomic
  transitions}},\ }\href {https://doi.org/10.1038/ncomms15501} {\bibfield
  {journal} {\bibinfo  {journal} {Nat. Commun.}\ }\textbf {\bibinfo {volume}
  {8}},\ \bibinfo {pages} {15501} (\bibinfo {year} {2017})}\BibitemShut
  {NoStop}%
\bibitem [{\citenamefont {He}\ \emph {et~al.}(2013)\citenamefont {He},
  \citenamefont {He}, \citenamefont {Wei}, \citenamefont {Wu}, \citenamefont
  {Atat{\"{u}}re}, \citenamefont {Schneider}, \citenamefont {H{\"{o}}fling},
  \citenamefont {Kamp}, \citenamefont {Lu},\ and\ \citenamefont
  {Pan}}]{He2013}%
  \BibitemOpen
  \bibfield  {author} {\bibinfo {author} {\bibfnamefont {Y.~M.}\ \bibnamefont
  {He}}, \bibinfo {author} {\bibfnamefont {Y.}~\bibnamefont {He}}, \bibinfo
  {author} {\bibfnamefont {Y.~J.}\ \bibnamefont {Wei}}, \bibinfo {author}
  {\bibfnamefont {D.}~\bibnamefont {Wu}}, \bibinfo {author} {\bibfnamefont
  {M.}~\bibnamefont {Atat{\"{u}}re}}, \bibinfo {author} {\bibfnamefont
  {C.}~\bibnamefont {Schneider}}, \bibinfo {author} {\bibfnamefont
  {S.}~\bibnamefont {H{\"{o}}fling}}, \bibinfo {author} {\bibfnamefont
  {M.}~\bibnamefont {Kamp}}, \bibinfo {author} {\bibfnamefont {C.~Y.}\
  \bibnamefont {Lu}},\ and\ \bibinfo {author} {\bibfnamefont {J.~W.}\
  \bibnamefont {Pan}},\ }\bibfield  {title} {\bibinfo {title} {{On-demand
  semiconductor single-photon source with near-unity indistinguishability}},\
  }\href {https://doi.org/10.1038/nnano.2012.262} {\bibfield  {journal}
  {\bibinfo  {journal} {Nat. Nanotechnol.}\ }\textbf {\bibinfo {volume} {8}},\
  \bibinfo {pages} {213} (\bibinfo {year} {2013})}\BibitemShut {NoStop}%
\bibitem [{\citenamefont {Braun}\ \emph {et~al.}(2016)\citenamefont {Braun},
  \citenamefont {Betzold}, \citenamefont {Lundt}, \citenamefont {Kamp},
  \citenamefont {H{\"{o}}fling},\ and\ \citenamefont {Schneider}}]{Braun2016}%
  \BibitemOpen
  \bibfield  {author} {\bibinfo {author} {\bibfnamefont {T.}~\bibnamefont
  {Braun}}, \bibinfo {author} {\bibfnamefont {S.}~\bibnamefont {Betzold}},
  \bibinfo {author} {\bibfnamefont {N.}~\bibnamefont {Lundt}}, \bibinfo
  {author} {\bibfnamefont {M.}~\bibnamefont {Kamp}}, \bibinfo {author}
  {\bibfnamefont {S.}~\bibnamefont {H{\"{o}}fling}},\ and\ \bibinfo {author}
  {\bibfnamefont {C.}~\bibnamefont {Schneider}},\ }\bibfield  {title} {\bibinfo
  {title} {{Impact of ex situ rapid thermal annealing on magneto-optical
  properties and oscillator strength of In(Ga)As quantum dots}},\ }\href
  {https://doi.org/10.1103/PhysRevB.93.155307} {\bibfield  {journal} {\bibinfo
  {journal} {Phys. Rev. B}\ }\textbf {\bibinfo {volume} {93}},\ \bibinfo
  {pages} {155307} (\bibinfo {year} {2016})}\BibitemShut {NoStop}%
\bibitem [{\citenamefont {Zhang}\ \emph {et~al.}(2015)\citenamefont {Zhang},
  \citenamefont {Wildmann}, \citenamefont {Ding}, \citenamefont {Trotta},
  \citenamefont {Huo}, \citenamefont {Zallo}, \citenamefont {Huber},
  \citenamefont {Rastelli},\ and\ \citenamefont {Schmidt}}]{zhang2015}%
  \BibitemOpen
  \bibfield  {author} {\bibinfo {author} {\bibfnamefont {J.}~\bibnamefont
  {Zhang}}, \bibinfo {author} {\bibfnamefont {J.~S.}\ \bibnamefont {Wildmann}},
  \bibinfo {author} {\bibfnamefont {F.}~\bibnamefont {Ding}}, \bibinfo {author}
  {\bibfnamefont {R.}~\bibnamefont {Trotta}}, \bibinfo {author} {\bibfnamefont
  {Y.}~\bibnamefont {Huo}}, \bibinfo {author} {\bibfnamefont {E.}~\bibnamefont
  {Zallo}}, \bibinfo {author} {\bibfnamefont {D.}~\bibnamefont {Huber}},
  \bibinfo {author} {\bibfnamefont {A.}~\bibnamefont {Rastelli}},\ and\
  \bibinfo {author} {\bibfnamefont {O.~G.}\ \bibnamefont {Schmidt}},\
  }\bibfield  {title} {\bibinfo {title} {High yield and ultrafast sources of
  electrically triggered entangled-photon pairs based on strain-tunable quantum
  dots},\ }\href@noop {} {\bibfield  {journal} {\bibinfo  {journal} {Nature
  communications}\ }\textbf {\bibinfo {volume} {6}},\ \bibinfo {pages} {1}
  (\bibinfo {year} {2015})}\BibitemShut {NoStop}%
\bibitem [{\citenamefont {Stockill}\ \emph {et~al.}(2016)\citenamefont
  {Stockill}, \citenamefont {{Le Gall}}, \citenamefont {Matthiesen},
  \citenamefont {Huthmacher}, \citenamefont {Clarke}, \citenamefont {Hugues},\
  and\ \citenamefont {Atat{\"{u}}re}}]{Stockill2016}%
  \BibitemOpen
  \bibfield  {author} {\bibinfo {author} {\bibfnamefont {R.}~\bibnamefont
  {Stockill}}, \bibinfo {author} {\bibfnamefont {C.}~\bibnamefont {{Le Gall}}},
  \bibinfo {author} {\bibfnamefont {C.}~\bibnamefont {Matthiesen}}, \bibinfo
  {author} {\bibfnamefont {L.}~\bibnamefont {Huthmacher}}, \bibinfo {author}
  {\bibfnamefont {E.}~\bibnamefont {Clarke}}, \bibinfo {author} {\bibfnamefont
  {M.}~\bibnamefont {Hugues}},\ and\ \bibinfo {author} {\bibfnamefont
  {M.}~\bibnamefont {Atat{\"{u}}re}},\ }\bibfield  {title} {\bibinfo {title}
  {{Quantum dot spin coherence governed by a strained nuclear environment}},\
  }\href {https://doi.org/10.1038/ncomms12745} {\bibfield  {journal} {\bibinfo
  {journal} {Nat. Commun.}\ }\textbf {\bibinfo {volume} {7}},\ \bibinfo {pages}
  {1} (\bibinfo {year} {2016})}\BibitemShut {NoStop}%
\bibitem [{\citenamefont {Gangloff}\ \emph {et~al.}(2019)\citenamefont
  {Gangloff}, \citenamefont {{\'{E}}thier-Majcher}, \citenamefont {Lang},
  \citenamefont {Denning}, \citenamefont {Bodey}, \citenamefont {Jackson},
  \citenamefont {Clarke}, \citenamefont {Hugues}, \citenamefont {{Le Gall}},\
  and\ \citenamefont {Atat{\"{u}}re}}]{Gangloff2019}%
  \BibitemOpen
  \bibfield  {author} {\bibinfo {author} {\bibfnamefont {D.~A.}\ \bibnamefont
  {Gangloff}}, \bibinfo {author} {\bibfnamefont {G.}~\bibnamefont
  {{\'{E}}thier-Majcher}}, \bibinfo {author} {\bibfnamefont {C.}~\bibnamefont
  {Lang}}, \bibinfo {author} {\bibfnamefont {E.~V.}\ \bibnamefont {Denning}},
  \bibinfo {author} {\bibfnamefont {J.~H.}\ \bibnamefont {Bodey}}, \bibinfo
  {author} {\bibfnamefont {D.~M.}\ \bibnamefont {Jackson}}, \bibinfo {author}
  {\bibfnamefont {E.}~\bibnamefont {Clarke}}, \bibinfo {author} {\bibfnamefont
  {M.}~\bibnamefont {Hugues}}, \bibinfo {author} {\bibfnamefont
  {C.}~\bibnamefont {{Le Gall}}},\ and\ \bibinfo {author} {\bibfnamefont
  {M.}~\bibnamefont {Atat{\"{u}}re}},\ }\bibfield  {title} {\bibinfo {title}
  {{Quantum interface of an electron and a nuclear ensemble}},\ }\href
  {https://doi.org/10.1126/science.aaw2906} {\bibfield  {journal} {\bibinfo
  {journal} {Science (80-. ).}\ }\textbf {\bibinfo {volume} {364}},\ \bibinfo
  {pages} {62} (\bibinfo {year} {2019})}\BibitemShut {NoStop}%
\bibitem [{\citenamefont {Zaporski}\ \emph {et~al.}(2023)\citenamefont
  {Zaporski}, \citenamefont {Shofer}, \citenamefont {Bodey}, \citenamefont
  {Manna}, \citenamefont {Gillard}, \citenamefont {Appel}, \citenamefont
  {Schimpf}, \citenamefont {{Covre da Silva}}, \citenamefont {Jarman},
  \citenamefont {Delamare}, \citenamefont {Park}, \citenamefont {Haeusler},
  \citenamefont {Chekhovich}, \citenamefont {Rastelli}, \citenamefont
  {Gangloff}, \citenamefont {Atat{\"{u}}re},\ and\ \citenamefont {{Le
  Gall}}}]{Zaporski2023}%
  \BibitemOpen
  \bibfield  {author} {\bibinfo {author} {\bibfnamefont {L.}~\bibnamefont
  {Zaporski}}, \bibinfo {author} {\bibfnamefont {N.}~\bibnamefont {Shofer}},
  \bibinfo {author} {\bibfnamefont {J.~H.}\ \bibnamefont {Bodey}}, \bibinfo
  {author} {\bibfnamefont {S.}~\bibnamefont {Manna}}, \bibinfo {author}
  {\bibfnamefont {G.}~\bibnamefont {Gillard}}, \bibinfo {author} {\bibfnamefont
  {M.~H.}\ \bibnamefont {Appel}}, \bibinfo {author} {\bibfnamefont
  {C.}~\bibnamefont {Schimpf}}, \bibinfo {author} {\bibfnamefont {S.~F.}\
  \bibnamefont {{Covre da Silva}}}, \bibinfo {author} {\bibfnamefont
  {J.}~\bibnamefont {Jarman}}, \bibinfo {author} {\bibfnamefont
  {G.}~\bibnamefont {Delamare}}, \bibinfo {author} {\bibfnamefont
  {G.}~\bibnamefont {Park}}, \bibinfo {author} {\bibfnamefont {U.}~\bibnamefont
  {Haeusler}}, \bibinfo {author} {\bibfnamefont {E.~A.}\ \bibnamefont
  {Chekhovich}}, \bibinfo {author} {\bibfnamefont {A.}~\bibnamefont
  {Rastelli}}, \bibinfo {author} {\bibfnamefont {D.~A.}\ \bibnamefont
  {Gangloff}}, \bibinfo {author} {\bibfnamefont {M.}~\bibnamefont
  {Atat{\"{u}}re}},\ and\ \bibinfo {author} {\bibfnamefont {C.}~\bibnamefont
  {{Le Gall}}},\ }\bibfield  {title} {\bibinfo {title} {{Ideal refocusing of an
  optically active spin qubit under strong hyperfine interactions}},\
  }\bibfield  {journal} {\bibinfo  {journal} {Nat. Nanotechnol.}\ }\href
  {https://doi.org/10.1038/s41565-022-01282-2} {10.1038/s41565-022-01282-2}
  (\bibinfo {year} {2023})\BibitemShut {NoStop}%
\bibitem [{\citenamefont {{Welander}}\ \emph {et~al.}(2014)\citenamefont
  {{Welander}}, \citenamefont {{Hildmann}},\ and\ \citenamefont
  {{Burkard}}}]{Welander2014}%
  \BibitemOpen
  \bibfield  {author} {\bibinfo {author} {\bibfnamefont {E.}~\bibnamefont
  {{Welander}}}, \bibinfo {author} {\bibfnamefont {J.}~\bibnamefont
  {{Hildmann}}},\ and\ \bibinfo {author} {\bibfnamefont {G.}~\bibnamefont
  {{Burkard}}},\ }\bibfield  {title} {\bibinfo {title} {{Influence of Hyperfine
  Interaction on the Entanglement of Photons Generated by Biexciton
  Recombination}},\ }\href@noop {} {\bibfield  {journal} {\bibinfo  {journal}
  {arXiv e-prints}\ ,\ \bibinfo {eid} {arXiv:1409.6521}} (\bibinfo {year}
  {2014})},\ \Eprint {https://arxiv.org/abs/1409.6521} {arXiv:1409.6521
  [cond-mat.mes-hall]} \BibitemShut {NoStop}%
\bibitem [{\citenamefont {Eble}\ \emph {et~al.}(2009)\citenamefont {Eble},
  \citenamefont {Testelin}, \citenamefont {Desfonds}, \citenamefont
  {Bernardot}, \citenamefont {Balocchi}, \citenamefont {Amand}, \citenamefont
  {Miard}, \citenamefont {Lema{\^{i}}tre}, \citenamefont {Marie},\ and\
  \citenamefont {Chamarro}}]{Eble2009}%
  \BibitemOpen
  \bibfield  {author} {\bibinfo {author} {\bibfnamefont {B.}~\bibnamefont
  {Eble}}, \bibinfo {author} {\bibfnamefont {C.}~\bibnamefont {Testelin}},
  \bibinfo {author} {\bibfnamefont {P.}~\bibnamefont {Desfonds}}, \bibinfo
  {author} {\bibfnamefont {F.}~\bibnamefont {Bernardot}}, \bibinfo {author}
  {\bibfnamefont {A.}~\bibnamefont {Balocchi}}, \bibinfo {author}
  {\bibfnamefont {T.}~\bibnamefont {Amand}}, \bibinfo {author} {\bibfnamefont
  {A.}~\bibnamefont {Miard}}, \bibinfo {author} {\bibfnamefont
  {A.}~\bibnamefont {Lema{\^{i}}tre}}, \bibinfo {author} {\bibfnamefont
  {X.}~\bibnamefont {Marie}},\ and\ \bibinfo {author} {\bibfnamefont
  {M.}~\bibnamefont {Chamarro}},\ }\bibfield  {title} {\bibinfo {title}
  {{Hole-nuclear spin interaction in quantum dots}},\ }\href
  {https://doi.org/10.1103/PhysRevLett.102.146601} {\bibfield  {journal}
  {\bibinfo  {journal} {Phys. Rev. Lett.}\ }\textbf {\bibinfo {volume} {102}},\
  \bibinfo {pages} {146601} (\bibinfo {year} {2009})}\BibitemShut {NoStop}%
\bibitem [{\citenamefont {{\'{E}}thier-Majcher}\ \emph
  {et~al.}(2017)\citenamefont {{\'{E}}thier-Majcher}, \citenamefont {Gangloff},
  \citenamefont {Stockill}, \citenamefont {Clarke}, \citenamefont {Hugues},
  \citenamefont {{Le Gall}},\ and\ \citenamefont
  {Atat{\"{u}}re}}]{Ethier-Majcher2017}%
  \BibitemOpen
  \bibfield  {author} {\bibinfo {author} {\bibfnamefont {G.}~\bibnamefont
  {{\'{E}}thier-Majcher}}, \bibinfo {author} {\bibfnamefont {D.}~\bibnamefont
  {Gangloff}}, \bibinfo {author} {\bibfnamefont {R.}~\bibnamefont {Stockill}},
  \bibinfo {author} {\bibfnamefont {E.}~\bibnamefont {Clarke}}, \bibinfo
  {author} {\bibfnamefont {M.}~\bibnamefont {Hugues}}, \bibinfo {author}
  {\bibfnamefont {C.}~\bibnamefont {{Le Gall}}},\ and\ \bibinfo {author}
  {\bibfnamefont {M.}~\bibnamefont {Atat{\"{u}}re}},\ }\bibfield  {title}
  {\bibinfo {title} {{Improving a Solid-State Qubit through an Engineered
  Mesoscopic Environment}},\ }\href
  {https://doi.org/10.1103/PhysRevLett.119.130503} {\bibfield  {journal}
  {\bibinfo  {journal} {Phys. Rev. Lett.}\ }\textbf {\bibinfo {volume} {119}},\
  \bibinfo {pages} {130503} (\bibinfo {year} {2017})}\BibitemShut {NoStop}%
\bibitem [{\citenamefont {Press}\ \emph {et~al.}(2010)\citenamefont {Press},
  \citenamefont {{De Greve}}, \citenamefont {McMahon}, \citenamefont {Ladd},
  \citenamefont {Friess}, \citenamefont {Schneider}, \citenamefont {Kamp},
  \citenamefont {H{\"{o}}fling}, \citenamefont {Forchel},\ and\ \citenamefont
  {Yamamoto}}]{Press2010}%
  \BibitemOpen
  \bibfield  {author} {\bibinfo {author} {\bibfnamefont {D.}~\bibnamefont
  {Press}}, \bibinfo {author} {\bibfnamefont {K.}~\bibnamefont {{De Greve}}},
  \bibinfo {author} {\bibfnamefont {P.~L.}\ \bibnamefont {McMahon}}, \bibinfo
  {author} {\bibfnamefont {T.~D.}\ \bibnamefont {Ladd}}, \bibinfo {author}
  {\bibfnamefont {B.}~\bibnamefont {Friess}}, \bibinfo {author} {\bibfnamefont
  {C.}~\bibnamefont {Schneider}}, \bibinfo {author} {\bibfnamefont
  {M.}~\bibnamefont {Kamp}}, \bibinfo {author} {\bibfnamefont {S.}~\bibnamefont
  {H{\"{o}}fling}}, \bibinfo {author} {\bibfnamefont {A.}~\bibnamefont
  {Forchel}},\ and\ \bibinfo {author} {\bibfnamefont {Y.}~\bibnamefont
  {Yamamoto}},\ }\bibfield  {title} {\bibinfo {title} {{Ultrafast optical spin
  echo in a single quantum dot}},\ }\href
  {https://doi.org/10.1038/nphoton.2010.83} {\bibfield  {journal} {\bibinfo
  {journal} {Nat. Photonics}\ }\textbf {\bibinfo {volume} {4}},\ \bibinfo
  {pages} {367} (\bibinfo {year} {2010})}\BibitemShut {NoStop}%
\bibitem [{\citenamefont {Stufler}\ \emph {et~al.}(2006)\citenamefont
  {Stufler}, \citenamefont {Machnikowski}, \citenamefont {Ester}, \citenamefont
  {Bichler}, \citenamefont {Axt}, \citenamefont {Kuhn},\ and\ \citenamefont
  {Zrenner}}]{Stufler2006}%
  \BibitemOpen
  \bibfield  {author} {\bibinfo {author} {\bibfnamefont {S.}~\bibnamefont
  {Stufler}}, \bibinfo {author} {\bibfnamefont {P.}~\bibnamefont
  {Machnikowski}}, \bibinfo {author} {\bibfnamefont {P.}~\bibnamefont {Ester}},
  \bibinfo {author} {\bibfnamefont {M.}~\bibnamefont {Bichler}}, \bibinfo
  {author} {\bibfnamefont {V.~M.}\ \bibnamefont {Axt}}, \bibinfo {author}
  {\bibfnamefont {T.}~\bibnamefont {Kuhn}},\ and\ \bibinfo {author}
  {\bibfnamefont {A.}~\bibnamefont {Zrenner}},\ }\bibfield  {title} {\bibinfo
  {title} {{Two-photon Rabi oscillations in a single Inx Ga1-x As GaAs quantum
  dot}},\ }\href {https://doi.org/10.1103/PhysRevB.73.125304} {\bibfield
  {journal} {\bibinfo  {journal} {Phys. Rev. B - Condens. Matter Mater. Phys.}\
  }\textbf {\bibinfo {volume} {73}},\ \bibinfo {pages} {125304} (\bibinfo
  {year} {2006})}\BibitemShut {NoStop}%
\bibitem [{\citenamefont {Huber}\ \emph {et~al.}(2019)\citenamefont {Huber},
  \citenamefont {Lehner}, \citenamefont {Csontosov{\'{a}}}, \citenamefont
  {Reindl}, \citenamefont {Schuler}, \citenamefont {{Covre Da Silva}},
  \citenamefont {Klenovsk{\'{y}}},\ and\ \citenamefont {Rastelli}}]{Huber2019}%
  \BibitemOpen
  \bibfield  {author} {\bibinfo {author} {\bibfnamefont {D.}~\bibnamefont
  {Huber}}, \bibinfo {author} {\bibfnamefont {B.~U.}\ \bibnamefont {Lehner}},
  \bibinfo {author} {\bibfnamefont {D.}~\bibnamefont {Csontosov{\'{a}}}},
  \bibinfo {author} {\bibfnamefont {M.}~\bibnamefont {Reindl}}, \bibinfo
  {author} {\bibfnamefont {S.}~\bibnamefont {Schuler}}, \bibinfo {author}
  {\bibfnamefont {S.~F.}\ \bibnamefont {{Covre Da Silva}}}, \bibinfo {author}
  {\bibfnamefont {P.}~\bibnamefont {Klenovsk{\'{y}}}},\ and\ \bibinfo {author}
  {\bibfnamefont {A.}~\bibnamefont {Rastelli}},\ }\bibfield  {title} {\bibinfo
  {title} {{Single-particle-picture breakdown in laterally weakly confining
  GaAs quantum dots}},\ }\href {https://doi.org/10.1103/PhysRevB.100.235425}
  {\bibfield  {journal} {\bibinfo  {journal} {Phys. Rev. B}\ }\textbf {\bibinfo
  {volume} {100}},\ \bibinfo {pages} {235425} (\bibinfo {year}
  {2019})}\BibitemShut {NoStop}%
\bibitem [{\citenamefont {Schimpf}\ \emph {et~al.}(2019)\citenamefont
  {Schimpf}, \citenamefont {Reindl}, \citenamefont {Klenovsk{\'{y}}},
  \citenamefont {Fromherz}, \citenamefont {{Covre Da Silva}}, \citenamefont
  {Hofer}, \citenamefont {Schneider}, \citenamefont {H{\"{o}}fling},
  \citenamefont {Trotta},\ and\ \citenamefont {Rastelli}}]{Schimpf2019}%
  \BibitemOpen
  \bibfield  {author} {\bibinfo {author} {\bibfnamefont {C.}~\bibnamefont
  {Schimpf}}, \bibinfo {author} {\bibfnamefont {M.}~\bibnamefont {Reindl}},
  \bibinfo {author} {\bibfnamefont {P.}~\bibnamefont {Klenovsk{\'{y}}}},
  \bibinfo {author} {\bibfnamefont {T.}~\bibnamefont {Fromherz}}, \bibinfo
  {author} {\bibfnamefont {S.~F.}\ \bibnamefont {{Covre Da Silva}}}, \bibinfo
  {author} {\bibfnamefont {J.}~\bibnamefont {Hofer}}, \bibinfo {author}
  {\bibfnamefont {C.}~\bibnamefont {Schneider}}, \bibinfo {author}
  {\bibfnamefont {S.}~\bibnamefont {H{\"{o}}fling}}, \bibinfo {author}
  {\bibfnamefont {R.}~\bibnamefont {Trotta}},\ and\ \bibinfo {author}
  {\bibfnamefont {A.}~\bibnamefont {Rastelli}},\ }\bibfield  {title} {\bibinfo
  {title} {{Resolving the temporal evolution of line broadening in single
  quantum emitters}},\ }\href {https://doi.org/10.1364/oe.27.035290} {\bibfield
   {journal} {\bibinfo  {journal} {Opt. Express}\ }\textbf {\bibinfo {volume}
  {27}},\ \bibinfo {pages} {35290} (\bibinfo {year} {2019})}\BibitemShut
  {NoStop}%
\bibitem [{\citenamefont {Bennett}\ \emph {et~al.}(2010)\citenamefont
  {Bennett}, \citenamefont {Pooley}, \citenamefont {Stevenson}, \citenamefont
  {Ward}, \citenamefont {Patel}, \citenamefont {de~La~Giroday}, \citenamefont
  {Sk{\"o}ld}, \citenamefont {Farrer}, \citenamefont {Nicoll}, \citenamefont
  {Ritchie} \emph {et~al.}}]{bennett2010}%
  \BibitemOpen
  \bibfield  {author} {\bibinfo {author} {\bibfnamefont {A.}~\bibnamefont
  {Bennett}}, \bibinfo {author} {\bibfnamefont {y.~M.}\ \bibnamefont {Pooley}},
  \bibinfo {author} {\bibfnamefont {n.~R.}\ \bibnamefont {Stevenson}}, \bibinfo
  {author} {\bibfnamefont {M.}~\bibnamefont {Ward}}, \bibinfo {author}
  {\bibfnamefont {R.}~\bibnamefont {Patel}}, \bibinfo {author} {\bibfnamefont
  {A.~B.}\ \bibnamefont {de~La~Giroday}}, \bibinfo {author} {\bibfnamefont
  {N.}~\bibnamefont {Sk{\"o}ld}}, \bibinfo {author} {\bibfnamefont
  {I.}~\bibnamefont {Farrer}}, \bibinfo {author} {\bibfnamefont
  {C.}~\bibnamefont {Nicoll}}, \bibinfo {author} {\bibfnamefont
  {D.}~\bibnamefont {Ritchie}}, \emph {et~al.},\ }\bibfield  {title} {\bibinfo
  {title} {Electric-field-induced coherent coupling of the exciton states in a
  single quantum dot},\ }\href@noop {} {\bibfield  {journal} {\bibinfo
  {journal} {Nature Physics}\ }\textbf {\bibinfo {volume} {6}},\ \bibinfo
  {pages} {947} (\bibinfo {year} {2010})}\BibitemShut {NoStop}%
\bibitem [{\citenamefont {Lettner}\ \emph {et~al.}(2021)\citenamefont
  {Lettner}, \citenamefont {Gyger}, \citenamefont {Zeuner}, \citenamefont
  {Schweickert}, \citenamefont {Steinhauer}, \citenamefont {{Reuterski{\"{o}}ld
  Hedlund}}, \citenamefont {Stroj}, \citenamefont {Rastelli}, \citenamefont
  {Hammar}, \citenamefont {Trotta}, \citenamefont {J{\"{o}}ns},\ and\
  \citenamefont {Zwiller}}]{lettner2021}%
  \BibitemOpen
  \bibfield  {author} {\bibinfo {author} {\bibfnamefont {T.}~\bibnamefont
  {Lettner}}, \bibinfo {author} {\bibfnamefont {S.}~\bibnamefont {Gyger}},
  \bibinfo {author} {\bibfnamefont {K.~D.}\ \bibnamefont {Zeuner}}, \bibinfo
  {author} {\bibfnamefont {L.}~\bibnamefont {Schweickert}}, \bibinfo {author}
  {\bibfnamefont {S.}~\bibnamefont {Steinhauer}}, \bibinfo {author}
  {\bibfnamefont {C.}~\bibnamefont {{Reuterski{\"{o}}ld Hedlund}}}, \bibinfo
  {author} {\bibfnamefont {S.}~\bibnamefont {Stroj}}, \bibinfo {author}
  {\bibfnamefont {A.}~\bibnamefont {Rastelli}}, \bibinfo {author}
  {\bibfnamefont {M.}~\bibnamefont {Hammar}}, \bibinfo {author} {\bibfnamefont
  {R.}~\bibnamefont {Trotta}}, \bibinfo {author} {\bibfnamefont {K.~D.}\
  \bibnamefont {J{\"{o}}ns}},\ and\ \bibinfo {author} {\bibfnamefont
  {V.}~\bibnamefont {Zwiller}},\ }\bibfield  {title} {\bibinfo {title}
  {{Strain-Controlled Quantum Dot Fine Structure for Entangled Photon
  Generation at 1550 nm}},\ }\href
  {https://doi.org/10.1021/acs.nanolett.1c04024} {\bibfield  {journal}
  {\bibinfo  {journal} {Nano Lett.}\ }\textbf {\bibinfo {volume} {21}},\
  \bibinfo {pages} {10501} (\bibinfo {year} {2021})}\BibitemShut {NoStop}%
\bibitem [{\citenamefont {James}\ \emph {et~al.}(2001)\citenamefont {James},
  \citenamefont {Kwiat}, \citenamefont {Munro},\ and\ \citenamefont
  {White}}]{James2001}%
  \BibitemOpen
  \bibfield  {author} {\bibinfo {author} {\bibfnamefont {D.~F.~V.}\
  \bibnamefont {James}}, \bibinfo {author} {\bibfnamefont {P.~G.}\ \bibnamefont
  {Kwiat}}, \bibinfo {author} {\bibfnamefont {W.~J.}\ \bibnamefont {Munro}},\
  and\ \bibinfo {author} {\bibfnamefont {A.~G.}\ \bibnamefont {White}},\
  }\bibfield  {title} {\bibinfo {title} {{Measurement of qubits}},\ }\href
  {https://doi.org/10.1103/PhysRevA.64.052312} {\bibfield  {journal} {\bibinfo
  {journal} {Phys. Rev. A}\ }\textbf {\bibinfo {volume} {64}},\ \bibinfo
  {pages} {052312} (\bibinfo {year} {2001})}\BibitemShut {NoStop}%
\bibitem [{\citenamefont {Wootters}(1998)}]{Wootters1998}%
  \BibitemOpen
  \bibfield  {author} {\bibinfo {author} {\bibfnamefont {W.~K.}\ \bibnamefont
  {Wootters}},\ }\bibfield  {title} {\bibinfo {title} {{Entanglement of
  Formation of an Arbitrary State of Two Qubits}},\ }\href
  {https://doi.org/10.1103/PhysRevLett.80.2245} {\bibfield  {journal} {\bibinfo
   {journal} {Phys. Rev. Lett.}\ }\textbf {\bibinfo {volume} {80}},\ \bibinfo
  {pages} {2245} (\bibinfo {year} {1998})}\BibitemShut {NoStop}%
\bibitem [{\citenamefont {Belhadj}\ \emph {et~al.}(2010)\citenamefont
  {Belhadj}, \citenamefont {Amand}, \citenamefont {Kunold}, \citenamefont
  {Simon}, \citenamefont {Kuroda}, \citenamefont {Abbarchi}, \citenamefont
  {Mano}, \citenamefont {Sakoda}, \citenamefont {Kunz}, \citenamefont {Marie},\
  and\ \citenamefont {Urbaszek}}]{Belhadj2010}%
  \BibitemOpen
  \bibfield  {author} {\bibinfo {author} {\bibfnamefont {T.}~\bibnamefont
  {Belhadj}}, \bibinfo {author} {\bibfnamefont {T.}~\bibnamefont {Amand}},
  \bibinfo {author} {\bibfnamefont {A.}~\bibnamefont {Kunold}}, \bibinfo
  {author} {\bibfnamefont {C.~M.}\ \bibnamefont {Simon}}, \bibinfo {author}
  {\bibfnamefont {T.}~\bibnamefont {Kuroda}}, \bibinfo {author} {\bibfnamefont
  {M.}~\bibnamefont {Abbarchi}}, \bibinfo {author} {\bibfnamefont
  {T.}~\bibnamefont {Mano}}, \bibinfo {author} {\bibfnamefont {K.}~\bibnamefont
  {Sakoda}}, \bibinfo {author} {\bibfnamefont {S.}~\bibnamefont {Kunz}},
  \bibinfo {author} {\bibfnamefont {X.}~\bibnamefont {Marie}},\ and\ \bibinfo
  {author} {\bibfnamefont {B.}~\bibnamefont {Urbaszek}},\ }\bibfield  {title}
  {\bibinfo {title} {{Impact of heavy hole-light hole coupling on optical
  selection rules in GaAs quantum dots}},\ }\href
  {https://doi.org/10.1063/1.3473824} {\bibfield  {journal} {\bibinfo
  {journal} {Appl. Phys. Lett.}\ }\textbf {\bibinfo {volume} {97}},\ \bibinfo
  {pages} {051111} (\bibinfo {year} {2010})}\BibitemShut {NoStop}%
\bibitem [{\citenamefont {Tonin}\ \emph {et~al.}(2012)\citenamefont {Tonin},
  \citenamefont {Hostein}, \citenamefont {Voliotis}, \citenamefont {Grousson},
  \citenamefont {Lemaitre},\ and\ \citenamefont {Martinez}}]{tonin2012}%
  \BibitemOpen
  \bibfield  {author} {\bibinfo {author} {\bibfnamefont {C.}~\bibnamefont
  {Tonin}}, \bibinfo {author} {\bibfnamefont {R.}~\bibnamefont {Hostein}},
  \bibinfo {author} {\bibfnamefont {V.}~\bibnamefont {Voliotis}}, \bibinfo
  {author} {\bibfnamefont {R.}~\bibnamefont {Grousson}}, \bibinfo {author}
  {\bibfnamefont {A.}~\bibnamefont {Lemaitre}},\ and\ \bibinfo {author}
  {\bibfnamefont {A.}~\bibnamefont {Martinez}},\ }\bibfield  {title} {\bibinfo
  {title} {Polarization properties of excitonic qubits in single self-assembled
  quantum dots},\ }\href@noop {} {\bibfield  {journal} {\bibinfo  {journal}
  {Physical Review B}\ }\textbf {\bibinfo {volume} {85}},\ \bibinfo {pages}
  {155303} (\bibinfo {year} {2012})}\BibitemShut {NoStop}%
\bibitem [{\citenamefont {Plumhof}\ \emph {et~al.}(2013)\citenamefont
  {Plumhof}, \citenamefont {Trotta}, \citenamefont {K\ifmmode~\check{r}\else
  \v{r}\fi{}\'apek}, \citenamefont {Zallo}, \citenamefont {Atkinson},
  \citenamefont {Kumar}, \citenamefont {Rastelli},\ and\ \citenamefont
  {Schmidt}}]{Plumhof2013}%
  \BibitemOpen
  \bibfield  {author} {\bibinfo {author} {\bibfnamefont {J.~D.}\ \bibnamefont
  {Plumhof}}, \bibinfo {author} {\bibfnamefont {R.}~\bibnamefont {Trotta}},
  \bibinfo {author} {\bibfnamefont {V.}~\bibnamefont {K\ifmmode~\check{r}\else
  \v{r}\fi{}\'apek}}, \bibinfo {author} {\bibfnamefont {E.}~\bibnamefont
  {Zallo}}, \bibinfo {author} {\bibfnamefont {P.}~\bibnamefont {Atkinson}},
  \bibinfo {author} {\bibfnamefont {S.}~\bibnamefont {Kumar}}, \bibinfo
  {author} {\bibfnamefont {A.}~\bibnamefont {Rastelli}},\ and\ \bibinfo
  {author} {\bibfnamefont {O.~G.}\ \bibnamefont {Schmidt}},\ }\bibfield
  {title} {\bibinfo {title} {Tuning of the valence band mixing of excitons
  confined in gaas/algaas quantum dots via piezoelectric-induced anisotropic
  strain},\ }\href {https://doi.org/10.1103/PhysRevB.87.075311} {\bibfield
  {journal} {\bibinfo  {journal} {Phys. Rev. B}\ }\textbf {\bibinfo {volume}
  {87}},\ \bibinfo {pages} {075311} (\bibinfo {year} {2013})}\BibitemShut
  {NoStop}%
\bibitem [{\citenamefont {Bulutay}(2012)}]{Bulutay2012}%
  \BibitemOpen
  \bibfield  {author} {\bibinfo {author} {\bibfnamefont {C.}~\bibnamefont
  {Bulutay}},\ }\bibfield  {title} {\bibinfo {title} {{Quadrupolar spectra of
  nuclear spins in strained In xGa 1-xAs quantum dots}},\ }\href
  {https://doi.org/10.1103/PhysRevB.85.115313} {\bibfield  {journal} {\bibinfo
  {journal} {Phys. Rev. B - Condens. Matter Mater. Phys.}\ }\textbf {\bibinfo
  {volume} {85}},\ \bibinfo {pages} {1} (\bibinfo {year} {2012})}\BibitemShut
  {NoStop}%
\bibitem [{\citenamefont {Denning}\ \emph {et~al.}(2019)\citenamefont
  {Denning}, \citenamefont {Gangloff}, \citenamefont {Atat{\"{u}}re},
  \citenamefont {M{\o}rk},\ and\ \citenamefont {{Le Gall}}}]{Denning2019}%
  \BibitemOpen
  \bibfield  {author} {\bibinfo {author} {\bibfnamefont {E.~V.}\ \bibnamefont
  {Denning}}, \bibinfo {author} {\bibfnamefont {D.~A.}\ \bibnamefont
  {Gangloff}}, \bibinfo {author} {\bibfnamefont {M.}~\bibnamefont
  {Atat{\"{u}}re}}, \bibinfo {author} {\bibfnamefont {J.}~\bibnamefont
  {M{\o}rk}},\ and\ \bibinfo {author} {\bibfnamefont {C.}~\bibnamefont {{Le
  Gall}}},\ }\bibfield  {title} {\bibinfo {title} {{Collective Quantum Memory
  Activated by a Driven Central Spin}},\ }\href
  {https://doi.org/10.1103/PhysRevLett.123.140502} {\bibfield  {journal}
  {\bibinfo  {journal} {Phys. Rev. Lett.}\ }\textbf {\bibinfo {volume} {123}},\
  \bibinfo {pages} {140502} (\bibinfo {year} {2019})}\BibitemShut {NoStop}%
\end{thebibliography}%

\end{document}